\documentclass[fleqn,usenatbib]{mnras}
\usepackage{siunitx}
\usepackage{mathptmx}

\usepackage[T1]{fontenc}

\DeclareRobustCommand{\VAN}[3]{#2}
\let\VANthebibliography\thebibliography
\def\thebibliography{\DeclareRobustCommand{\VAN}[3]{##3}\VANthebibliography}

\usepackage{graphicx}	
\usepackage{amsmath}	
\usepackage{amssymb}	

\title[{\textsc{thesan-hr}}]{{\textsc{thesan-hr}}: How does reionization impact early galaxy evolution?}

\author[Josh Borrow et al.]{Josh Borrow,$^{1}$\thanks{E-mail: josh@joshborrow.com, borrowj@mit.edu (JB)}
Rahul~Kannan,$^{2}$
Enrico Garaldi,$^{3}$
Aaron~Smith,$^{2, 1}$
Mark~Vogelsberger,$^{1,4}$
\newauthor
R\"{u}diger~Pakmor,$^{3}$
Volker~Springel,$^{3}$
and Lars~Hernquist$^{2}$
\\
\\
$^{1}$Department of Physics and Kavli Institute for Astrophysics and Space Research, Massachusetts Institute of Technology, Cambridge, MA 02139, USA\\
$^{2}$Center for Astrophysics $\vert$ Harvard \& Smithsonian, 60 Garden Street, Cambridge, MA 02138, USA \\
$^{3}$Max-Planck Institute for Astrophysics, Karl-Schwarzschild-Str.~1, D-85741 Garching, Germany \\
$^{4}$The NSF AI Institute for Artificial Intelligence and Fundamental Interactions, Massachusetts Institute of Technology, Cambridge MA 02139, USA
}

\date{Accepted XXX. Received YYY; in original form ZZZ}

\pubyear{2022}

\begin{document}
\label{firstpage}
\pagerange{\pageref{firstpage}--\pageref{lastpage}}
\maketitle

\begin{abstract}	
The feedback loop between the galaxies producing the background radiation field
for reionization and their growth is crucial, particularly for low-mass haloes.
Despite this, the vast majority of galaxy formation studies employ a
spatially-uniform, time-varying reionizing background, with the majority of reionization
studies employing galaxy formation models only required to work at high
redshift. This paper uses the well-studied TNG galaxy formation model,
calibrated at low redshift, coupled to the Arepo-RT code, to self-consistently
solve the coupled problems of galaxy evolution and reionization, evaluating the
impact of patchy (and slow) reionization on early galaxies. {\textsc{thesan-hr}}
is an extension of the {\textsc{thesan}} project to higher resolution (a factor
of 50 increase, with a baryonic mass of $m_{\rm b} \approx 10^4$~M$_\odot$), to
additionally enable the study of `mini-haloes' with virial temperatures $T_{\rm
vir} < 10^4$~K.  Comparing the self-consistent model to a uniform UV background,
we show that galaxies in {\textsc{thesan-hr}} are predicted to be larger in
physical extent (by a factor $\sim 2$), less metal enriched (by $\sim 0.2$~dex),
and less abundant (by a factor $\sim 10$ at $M_{\rm 1500}~=~-10$) by $z=5$. We
show that differences in star formation and enrichment patterns lead to
significantly different predictions for star formation in low mass haloes,
low-metallicity star formation, and even the occupation fraction of haloes. We
posit that cosmological galaxy formation simulations aiming to study early
galaxy formation ($z \gtrsim 3$) must employ a spatially inhomogeneous UV
background to accurately reproduce galaxy properties.
\end{abstract}

\begin{keywords}
Galaxies: formation;
Galaxies: evolution;
Cosmology: dark ages, reionization, first stars;
Methods: numerical
\end{keywords}



\section{Introduction}
    
Just after the Big Bang, the Universe was filled with a hot, dense, plasma,
where radiation and matter were in thermal equilibrium equilibrium. As the
Universe expanded and cooled, electrons and protons could recombine, filling the
Universe with a primordial mixture of neutral hydrogen and helium. In this early
state, small overdensities began to grow under their own self-gravity, beginning
the process of cosmological structure formation. As gas accreted into dark
matter haloes, cooling down to temperatures low enough to form molecular
hydrogen, the first stars were born. These massive, hot, and bright Population
(Pop) III stars produced copious amounts of ionizing radiation, which began the
process of reionizing the Universe. As the abundance of stars in galaxies grew,
the most massive galaxies were able to reionize the Universe, fully ionizing
the intergalactic medium (IGM), leaving it in equilibrium with the background
radiation field once again, at a temperature of $T \sim 10^4$ K \citep[see][for
an extensive review of reionization]{Wise2019, Gnedin2022}.

Understanding the process of reionization is central to our understanding
of the evolution of the early Universe, but at the core of the common description
(as above) is a circular argument -- galaxies are made out of cold, neutral,
gas, but they destroy and suppress the formation of such gas. In addition, as ionization
fronts proceed outwards from galaxies, and can only travel as quickly as they can
ionize gas, reionization is expected to be patchy, rather than spatially
uniform. It is well understood how such patchy reionization models can impact
the filamentary structures between galaxies and the IGM more generally,
with areas that are reionized earlier having lower densities due to additional
photoheating pressure \citep{Gnedin1998, Theuns2000, Schaye2000, Kulkarni2015,
Rorai2017, Wu2019, Katz2020, Puchwein2022}. Strong radiation fields are also
expected to reduce the fraction of baryons that galaxies can retain, further
suppressing galaxy growth, as well as influencing the makeup of the multiphase
interstellar medium \citep{Wise2008, Wise2012b, Wu2019, Katz2020}.

Despite this, most cosmological galaxy formation simulations typically employ a
time varying, but spatially-uniform, ultraviolet background (UVB) to model the
photoionizing feedback from their constituent stars. These models are entirely
fixed, and not self-consistent with the constituents of the simulation
\citep{Haardt1996, Haardt2001,Faucher-Giguere2009, Dubois2012, Vogelsberger2013,
Schaye2015, Ploeckinger2020}. Typically, this approach adopts a fixed `reionization
redshift', often $z_{\rm reion}\approx 10$, where all gas in the simulation
is exposed to a strong UV background following these time-varying models,
whereas at prior times the gas is considered to be shielded. This
approximation is certainly valid at present times, where all gas will have been
exposed to such radiation fields for many billions of years, when typical galaxy
formation models are employed \citep[the EAGLE model forms 90\% of its stellar
mass in the time between $z=2$ and $z=0$;][]{Crain2015}.  During (and shortly
after) the first epoch of galaxy formation, however, such approximations are not
likely to be valid. As we enter the epoch of the \emph{JWST}, and galaxy
formation studies move to redshifts $z > 3$ (with galaxy surveys such as
\emph{JADES}), we must re-examine the impact of reionization on our galaxy
formation models \citep{Kannan2022b}.

There have been many studies that attempt to model reionization consistently
with galaxy formation simulations by post-processing the outputs, using the gas
density field and stellar masses along with either an approximate method
\citep[e.g. excursion set formalism, dark matter, or gas;][]{Santos2010,
Mesinger2011, Hassan2016, Whitler2020} or a radiative transfer (RT) calculation
\citep{Baek2010, Ciardi2012, Graziani2015,Ross2017, Eide2018, Hassan2022} to
estimate the speed of ionization fronts and morphologies of the ionized bubbles
in the early universe.  This approach is sensitive to many factors, not least
the escape fractions of photons from galaxies which are still relatively
unconstrained \citep{Kostyuk2022,Yeh2022,Rosdahl2022}, or the impact of dust on
galaxy luminosity predictions \citep{Vogelsberger2020c, Shen2020, Shen2022}, but
can typically perform RT calculations at high accuracy (with many frequency
bins), and in much larger volumes than those that use fully-coupled RT to
\emph{self-consistently} study reionization.

Fully self-consistent simulations of reionization require on-the-fly coupling of
RT, hydrodynamics, and cooling calculations. Because of the requirement to
effectively sample the \emph{light}-crossing timescale (rather than the
sound-crossing timescale required for hydrodynamics calculations), such models
are extremely computationally intensive, an order of magnitude more computing
time than hydrodynamical calculations, even when employing a reduced speed of
light approximation \citep{Kannan2022}. Despite this additional cost, such
simulations are extremely fruitful, and there are a handful of models available
today for use by the community.  These schemes either target a self-consistent
description of the largest scales with boxes $L_{\rm box} \gtrapprox 100$ Mpc at
relatively low resolution \citep{O'Shea2015, Ocvirk2016, Ocvirk2020}, with
recent models increasing in resolution significantly to track even the formation
of dwarfs \citep{Kannan2022, Smith2022, Garaldi2022, Rosdahl2022}, or aim to
target the formation of individual dwarf galaxies or small volumes
\citep{Petkova2011, Gnedin2014, So2014, Gnedin2016, Pawlik2017,
Semelin2017,Trebitsch2017, Rosdahl2018, Pallottini2019, Obreja2019, Wu2019,
Pallottini2022}. In this paper, we aim to extend one of these large-volume
suites, {\textsc{thesan}} \citep{Kannan2022, Smith2022, Garaldi2022}, down to
even lower masses and higher resolutions (with a resolution increase of a factor
100) using the same model to span halo mass scales of $10^6 < M_{\rm H} / {\rm
M}_\odot < 10^{12}$ at $z=6$.

Simulating to low mass scales is crucial to fully understand the 
galaxy-reionization connection. Haloes with a virial temperature $T_{\rm vir} <
10^4$ K (circular velocities $V_{c} < 30$ km s$^{-1}$, M$_{\rm H} \approx 10^8$
M$_\odot$ at $z=5$, also known as `minihaloes') are susceptible to
photoevaporation from the UV background, as photoheated {\textsc{Hii}} gas in ionization
equilibrium is typically around this temperature \citep{Rees1986, Shapiro2004,
Iliev2005}. As such, galaxy formation simulations with a resolution too low to
model these haloes will miss out on a population of galaxies that are highly
impacted by reionization physics. Other consequences of photoheating, such as
prolonged cooling times, or preventative feedback, are less affected by the
specific choice of mass scale, though clearly will be more efficient in
lower-mass objects \citep{Efstathiou1992,Shapiro1994, Gnedin2000, Hoeft2006,
Okamoto2008, Gnedin2014b, Katz2020}. These haloes are additionally influenced by
streaming velocities between gas and dark matter, though this is expected to be
a subdominant effect \citep{Milosavljevic2014, Schauer2022}.

To fully understand the impact of our choice of reionization treatment, we perform
simulations using {\tt{Arepo-RT}} \citep{Kannan2019} and the fiducial
{\textsc{thesan}} model, as used in the {\textsc{thesan-1}} simulation volume.
We choose to compare two scenarios: the typical \citet{Faucher-Giguere2009}
time-varying, but spatially-uniform UVB used in most cosmological galaxy
formation simulations, and the fully self-consistent model from
{\textsc{thesan}}, to investigate what physics simulations employing a spatially
uniform UVB are missing. These two scenarios also can correspond to two
different fields: one near a massive, highly ionizing object (the uniform UVB),
and one without, where the radiation from low mass sources dominates (our
{\textsc{thesan}} model). 

The paper is organised as follows: in
section \ref{sec:simulations}, we describe our simulation model, choice of
simulation volume, and some global properties of our volumes over time. To
demonstrate in detail the impact of reionization treatments on cumulative galaxy
properties, such as the UV luminosity function, we show various scaling
relations in section \ref{sec:galaxies}. In section \ref{sec:gas}, we tie these
differing galaxy properties to different gas phase evolution between the
reionization treatments, and in section \ref{sec:conclusions} we summarise and
conclude.

 

\section{Simulations}
\label{sec:simulations}

In this paper we explore two classes of models: simulations with a typical,
spatially-uniform, but time-varying UV background, and simulations that include
a self-consistent treatment using the same on-the-fly radiative transfer code as
in the main {\textsc{thesan}} project. We perform simulations with the exact
same initial conditions, the same underlying physical model (including
nonequlibrium chemistry), and the same output strategy, in all cases, with the
only difference being our choice of reionization treatment (here referred to as
our `reionization model'). In this section we describe both the model and
simulation volumes that we will use throughout the paper.

\begin{figure*}
    \centering
    \includegraphics{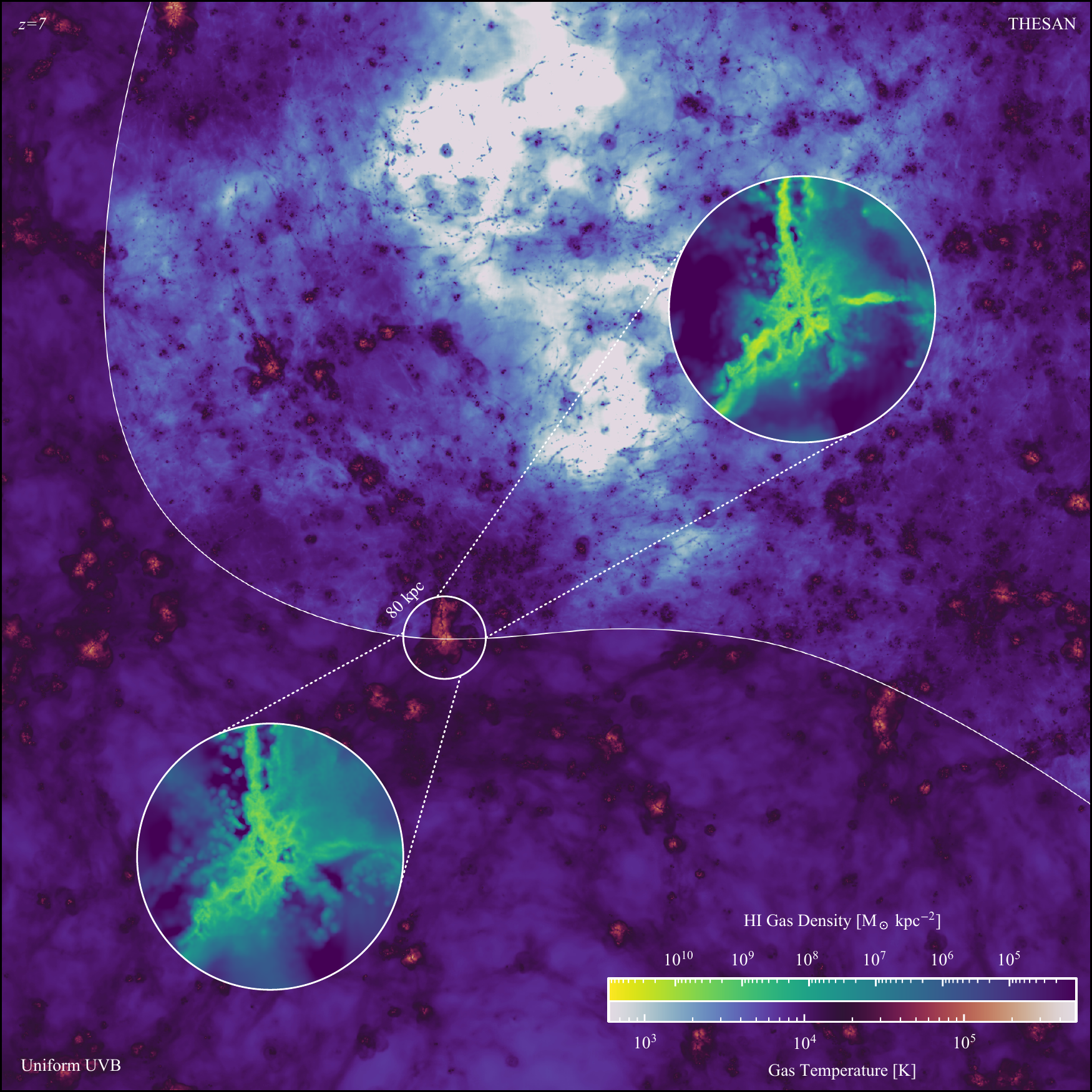}
    \caption{A projection through the entire L8N512 volume, at redshift $z=7$,
    showing the mass-weighted temperature along the line of sight (background).
    Above the curved white line is the {\textsc{thesan}} simulation, and below the line
    is the Uniform UVB simulation. The {\textsc{thesan}} simulation shows
    significantly cooler temperatures in void regions, with these areas
    remaining neutral. The circle, which is 80 physical kpc in diameter,
    surrounds the most massive galaxy in the volume, with zooms shown for both
    simulations. Within the zooms, the {\textsc{Hi}} gas density is shown,
    with the {\textsc{thesan}} simulation exhibiting significantly ($\approx 10\times$) higher
    {\textsc{Hi}} fractions than the Uniform UVB one.}
    \label{fig:bigplot}
\end{figure*}

In Fig. \ref{fig:bigplot} we show the large-scale gas temperature distribution
at redshift $z=7$ for these two models. The top right part of the image shows
the {\textsc{thesan-hr}} volume, which contains much colder regions with
temperatures $T < 10^3$ K that are not possible in the Uniform UVB simulation
due to the uniform photoheating of the gas. This leads additionally to there
being much denser (and spatially thinner) cold filaments in {\textsc{thesan}},
though the hot collisionally ionized regions that are heated by stellar feedback
remain roughly consistent between the two models. This overall picture is
consistent with prior work on patchy reionization, e.g. \citet{Puchwein2022}.

In the circles, we show the {\textsc{Hi}} gas in the most massive galaxy (halo mass $M_{\rm
H} \approx 5 \times 10^{10}$ M$_\odot$, stellar mass $M_* \approx 1 \times 10^8$
M$_\odot$) in the volume. In the Uniform UVB model this gas is produced through
self-shielding using the prescription from \citet{Rahmati2014}. In
{\textsc{thesan}}, self-shielding is calculated self-consistently from the
radiation fields and recombination rates of the ionized gas.  This leads to
differing {\textsc{Hi}} gas content, with {\textsc{thesan}} showing a factor of $\sim 10$
greater {\textsc{Hi}} gas density in filaments, with a factor $\sim 10$ lower density in
the surrounding CGM.

In the remainder of the paper, we aim to understand the impacts of these two
reionization models, and hence different gas phase structures, on the process of
galaxy formation by studying in detail the properties of galaxies.

\subsection{Simulation Code}

The simulations in this paper were performed using {\textsc{
Arepo-RT}} \citep{Kannan2019}, which is based upon the {\textsc{Arepo}} moving mesh
magnetohydrodynamics code \citep{Springel2011, Pakmor2013, Weinberger2020}.  They are an
extension of the {\textsc{thesan}} project \citep{Kannan2022, Smith2022, Garaldi2022}, and
use the same physical and numerical models as these simulations. We briefly summarise
these models here, and refer interested readers to Section 2 of \citet{Kannan2022}
for a detailed descrpition.

{\textsc{Arepo-RT}} solves the (radiation-, magneto-) hydrodynamical equations on an 
unstructured Voronoi mesh, with gas `particles' acting as mesh-generating points
\citep{Pakmor2013}. This moving mesh solver is quasi-Lagrangian, and solves the
hydrodynamic equations in the rest frame of an interface between all mesh
generating points, which are regularized frequently \citep{Vogelsberger2012,
Pakmor2016}. Gravitational forces are solved (including periodicity) using a
Hybrid Tree-PM approach which employs an oct-tree to solve for short-range forces,
with long range forces computed with the particle mesh method
\citep{Springel2005b, Springel2021}.

Radiation transport is handled using a moment-based approach, and solves the
hyperbolic conservation equations for photon number density and flux using the
M1 closure scheme \citep{Levermore1984, Dubroca1999}. The UV continuum is
grouped into three frequency bins with energy intervals $[13.6, 24.6, 54.4,
\infty)$ eV, and we use the reduced speed of light $\tilde{c} = 0.2c$ and
subcycling (allowing 32 subcycles) of the radiation transport to improve the
speed of the calculation. We couple these photons to the gas using a
non-equilibrium thermochemistry module that tracks abundances of five species:
HI, {\textsc{Hii}}, HeI, HeII, and HeIII. The final cooling rates for gas
include contributions from this non-equilibrium thermochemistry (i.e. primordial
cooling), equilibrium models for metal cooling following
\citet{Vogelsberger2013}, and compton cooling. Our metal cooling rates are
computed assuming metals are in equilibrium with a spatially-uniform UV
background with the radiation field given by a \citet{Faucher-Giguere2009}
model, which is potentially problematic given that reionization proceeds
differently in our simulations. Unfortunately a full calculation including
nonequlibrium metal cooling on the scales studied here is infeasible given
present computational restraints, and we are mainly interested in differences
between our two reionization models rather than making absolute predictions. 

To simulate key processes that occur on scales smaller than those resolved in the
simulation, notably star formation, stellar feedback, black hole formation,
and AGN feedback, we use the IllustrisTNG galaxy formation model
\citep{Marinacci2018, Naiman2018, Nelson2018, Pillepich2018b, Springel2018,Pillepich2019,Nelson2019b,Nelson2019},
an update to the previous Illustris galaxy formation model \citep{Vogelsberger2014}.
We use the sub-resolution treatment of the interstellar medium (ISM) from
\citep{Springel2003}, star formation and feedback following \citep{Pillepich2018},
and AGN formation and accretion following \citep{Weinberger2017}. In addition,
we use the empirical dust treatments from \citet{McKinnon2016, McKinnon2017}, 
and track dust properties of each gas cell.

Radiation is produced by stars and AGN, though in the volumes studied here the
contribution from AGN is negligible as there are only four black holes in our
largest volume, with a maximum mass of $M_{\rm BH} \approx 5 \times 10^6$
M$_\odot$. As such, we refer the reader to \citet{Kannan2022} for a description
of our black hole and AGN implementation, and only describe the implementation 
of stars and their feedback here. The luminosity and spectral energy density of
stars in {\textsc{thesan}} is given as a complex function of age and metallicity taken from
the Binary Population and Spectral Synthesis models \citep[BPASS
v2.2.1][]{Eldridge2017}. The sub-grid escape fraction of stars was set to be
$f_{\rm esc} = 0.37$ in the original {\textsc{thesan}} simulations to match the
global reionization history of the Universe, and we adopt the same value here for
consistency.

For the models employing a Uniform UV Background we assume reionization is a
near-instantaneous process occurring at redshift $z=10$ (following the original
IllustrisTNG model\footnote{We note that the original Illustris and IllustrisTNG
simulations used a fixed reionization redshift of $z=6$, though similar models
like EAGLE initiate their UVB at an even higher redshift of $z=11.5$
\citep{Schaye2015}.  We choose here to use $z=10$ as our fully coupled radiation
transport simulations end at $z=5$ to ensure efficient use of computational
resources.}) with the strength of the radiation field given by the prescription
from \citet{Faucher-Giguere2009}, as this is a standard practice in cosmological
galaxy formation simulations. Simulations with this UVB are performed in the
exact same way as the {\textsc{thesan}} models, but all sources of radiation
have their escape fraction set to $f_{\rm esc} = 0$. The ionizing radiation flux
from the UVB is then passed directly to the thermochemistry module, in most
cases fully ionizing the gas, though gas can be self-shielded from such
radiation if it reaches a high enough density. We employ the model from
\citet{Rahmati2014} to determine if such gas is self-shielded against the
external UVB.

We note here that these new simulations using a Uniform UV background are
different than those performed during the original {\textsc{thesan}} simulations, 
for instance {\textsc{thesan-tng-2}}. The prior simulations that used a spatially
uniform reionization history and disabled the nonequlibrium thermochemistry module
entirely, which leads to small, but systematic, changes in galaxy properties 
\citep[see][Appendix A]{Garaldi2022}.

Our simulation models then hence contain all of the typically included physics
for a standard galaxy formation simulation (i.e. those targeting $z=0$, similar
to those employed in, for instance, EAGLE, IllustrisTNG, or SIMBA), with the
addition of the required physics for reionization (radiative transfer, photon
production, and non-equlibrium hydrogen and helium ionisation). We do not
include the physics required to accurately track the formation of the first
stars (for instance, molecular hydrogen formation, Lyman-Werner radiation, or
explicit Population-III star modelling) and the dynamics within mini-haloes.
Because of this, we refrain from a full investigation of the initial star
formation events, and focus mainly on the differences seen in galaxy populations
due to the varying reionization schemes as they pertain to the accuracy of the
galaxy formation models.

\subsection{Simulation Volumes}

\begin{table*}
    \centering
    \caption{Setup information about the two volumes used in this study, along
    with the main {\textsc{thesan}}-1 volume, with information (from left to
    right) giving: the name of this box within the paper, the total
    number of particles within the volume, the co-moving box volume, the initial
    mean mass of gas particles within the simulation, the mass of dark matter
    particles, and the constant physical gravitational softening used for
    baryons and dark matter respectively.}
    \begin{tabular}{c c c c c c c}
        \hline
        Name & Number of Particles & Box Length [cMpc] & $m_{\rm g}$ [M$_\odot$] & $m_{\rm DM}$ [M$_\odot$] & $\epsilon_{\rm g}$ [ckpc] & $\epsilon_{\rm DM}$ [ckpc] \\
        \hline
        {\textsc{thesan}}-1 & $2 \times 2100^3$ & 95.5 & $5.82 \times 10^5$ & $3.12 \times 10^6$ & 2.2 & 2.2 \\
        L4N512 & $2 \times 512^3$ & 5.9 & $1.13\times10^{4}$ & $6.03 \times 10^{4}$ & 0.425 & 0.425 \\
        L8N512 & $2 \times 512^3$ & 11.8 & $9.04 \times 10^{4}$ & $4.82 \times 10^{5}$ & 0.85 & 0.85 \\
        \hline
    \end{tabular}
    \label{tab:res}
\end{table*}

Our simulation volumes are generated following a typical
\citet{PlanckCollaboration2016} cosmology, with $H_0 = 100h$, $h=0.6774$,
$\Omega_{\rm m} =0.3089$, $\Omega_\Lambda = 0.6911$, $\Omega_{\rm b} = 0.0486$,
$\sigma_8=0.8159$, and $n_{\rm s} = 0.9667$, with all symbols taking their usual
meanings. Gas is assumed to have initially primordial composition with hydrogen
mass fraction $X = 0.76$ and helium mass fraction $Y = 1 - X$.

Throughout this paper we use two main simulation boxes, which have co-moving
volumes $(4 / h)^3$ cMpc$^3$ and $(8 / h)^3$ cMpc$^3$, named L4N512 and L8N512
respectively.  Both volumes contain $512^3$ dark matter particles and gas cells
initially, giving different mass resolutions (and hence softenings, etc.) for
both volumes, as described in Table \ref{tab:res}. We note here that all further
quantities within this paper are provided $h$-free, aside from the labels for
boxsize denomination.  These two volumes were evolved with the two reionization
models, a uniform UVB, and the full {\textsc{thesan}} model, for four total
simulations considered in this paper.

Our simulation volumes are chosen to be small for computational efficiency reasons.
Simulating, with radiative transfer, at these resolutions is computationally
challenging. Our small volumes hence come with the following important caveats:

\begin{enumerate}
\item The small volumes are at the mean density of the universe, which on average should
be partially reionized by massive external sources. Our volumes, however, will
always reionize internally, due to the lack of any external sources. As such,
we should expect our volumes to reionize slower on average than a fully representative
volume, with the smaller $(4 / h)^3$~Mpc$^{3}$ volumes reionizing on longer timescales.

\item Due to the resolution differences between the volumes, and the cold dark matter
model employed, we will see significant differences in the populations of
low-mass haloes (i.e. there will be significantly more in the higher resolution,
smaller volume, case). As these low-mass haloes may change the topology of
reionization, we may not see strong convergence between resolution levels until
haloes much more massive than the resolution limit (e.g. $> 1000$ particles, rather
than the typical $32-100$ particle limit).
\end{enumerate}

\begin{figure}
    \centering
    \includegraphics{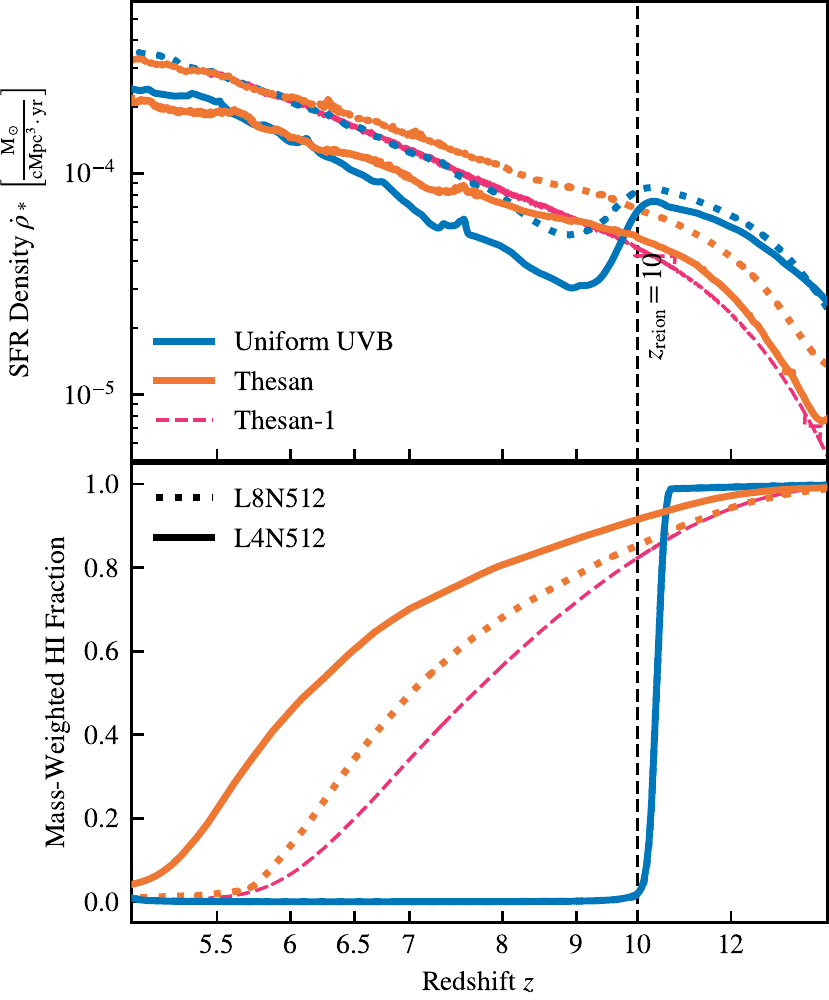}
    \caption{The star formation rate density history of both simulations (top)
    and the global mass-weighted {\textsc{Hi}} fraction history (bottom). Indicated by the
    black dashed line is the instantaneous reionization redshift of $z=10$ in the
    Uniform UVB model. Pre-reionization, the Uniform UVB model is able to
    sustain a higher global star formation rate density, due to a lack of
    photoionization heating from stars. Post-reionization, this star formation
    rate drops significantly (a factor of $\approx 4$) as star forming regions in
    low-mass galaxies are evaporated by the strong UV background. The {\textsc{thesan}}
    model reionizes much less quickly, fully reionizing the volume (i.e. a HI
    fraction of approximately zero) around redshift $z=5.6$. In pink, we show
    the reionization history of the original {\textsc{thesan-1}} simulation for comparison.}
    \label{fig:reionizationhistory}
\end{figure}

In the top panel of Fig. \ref{fig:reionizationhistory} we show the star
formation history of our four simulations alongside the {\textsc{thesan-1}}
simulation.  {\textsc{thesan-1}} matches well with both of our {\textsc{thesan}}
volumes, with there being some small offsets in the star formation rate density
between models (roughly $\sim 0.1$ dex, notably not monotonic with resolution;
we show in Appendix \ref{app:res} the resolution convergence of our results,
with the differences here mainly driven by the volume, not resolution).  By
contrast, the Uniform UVB models show significantly higher ($\sim 0.3$ dex) star
formation rates than {\textsc{thesan}} before reionization ($z > 10$) that are
rapidly damped post-reionization.  Our L4N512 volume with the Uniform UVB has a
$\sim 0.3$ dex lower star formation rate than its {\textsc{thesan}} counterpart
at $z=9$, shortly after reionization, but by $z=5.5$, the two reionization
models match (aside from differences between volumes). This provides early
evidence that the {\textsc{thesan}} and Uniform UVB models have broadly
acceptable levels of strong convergence at the resolutions studied here ($\sim
10^6$ M$_\odot$ for baryons in {\textsc{thesan-1}}, to $\sim 10^4$ M$_\odot$ for
our highest resolution {\textsc{thesan}} run).

\begin{figure*}
    \centering
    \includegraphics{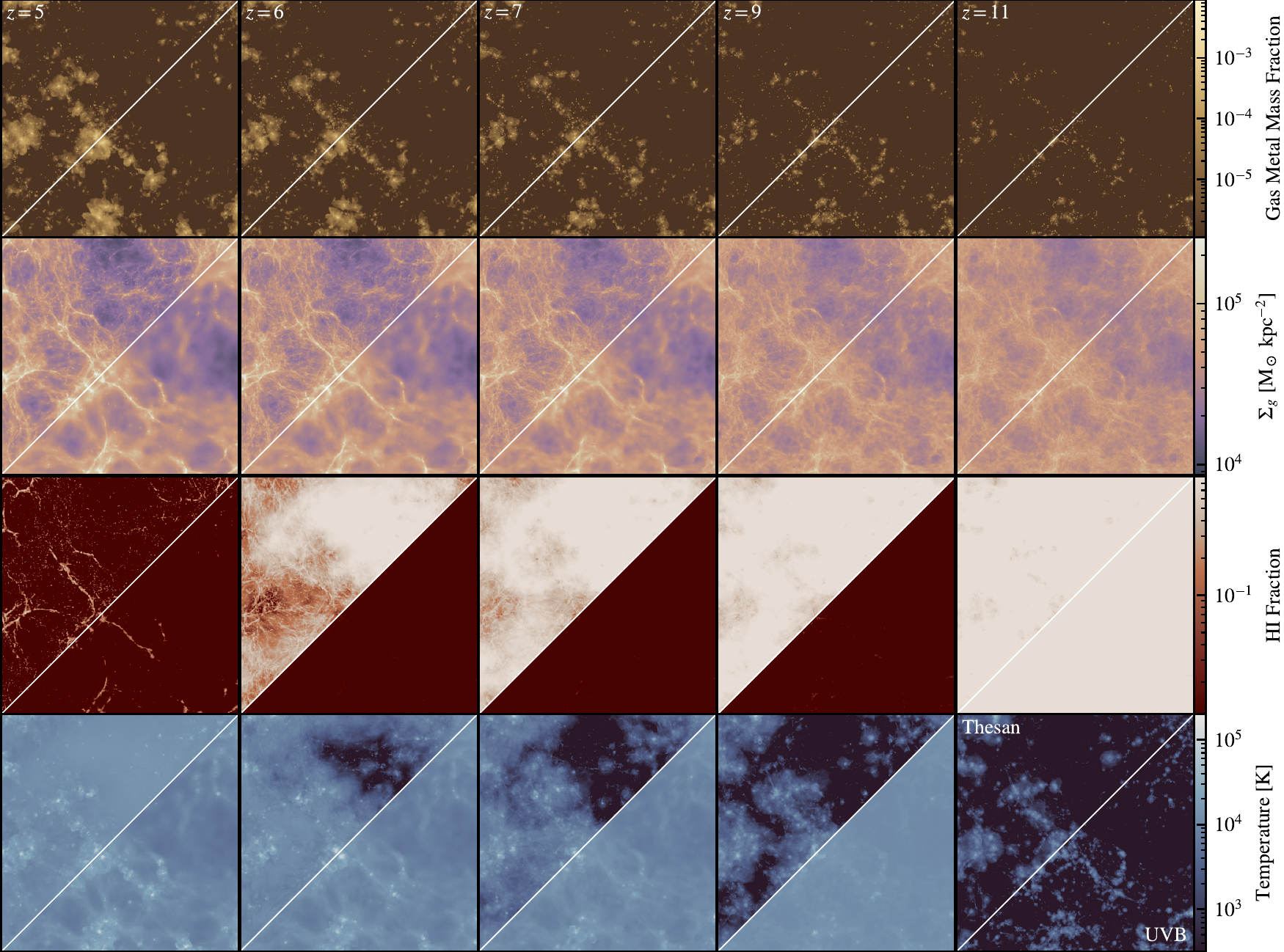}
    \caption{Macroscopic evolution of four key quantities in the L4N512 volume
    (with the entire volume shown in projection; quantities are calculated as
    the mass-weighted mean over the  line of sight): the gas metal mass
    fraction, defined as the fraction of mass of each gas cell in the metal
    phase, the projected gas density $\Sigma_{\rm g}$, the {\textsc{Hi}} gas fraction, and
    the temperature. The lower right of each panel shows the result from the
    Uniform UVB simulation, with the upper left showing the same part of the
    volume (mirrored) but in the {\textsc{thesan}} simulation.}
    \label{fig:combinedboxes}
\end{figure*}

The bottom panel of Fig. \ref{fig:reionizationhistory} shows the mass-weighted
{\textsc{Hi}} fraction as a function of redshift. The mass-weighted {\textsc{Hi}} fraction is the
total mass of {\textsc{Hi}} gas in the volume divided by the total gas mass in the volume
at that epoch. Here we see small differences in the reionization history between
the {\textsc{thesan}} volumes; as we decrease box-size, we see that the volumes
reionize more slowly, with L4N512 reionizing the slowest (at $z=5.5$, it still
has an {\textsc{Hi}} fraction of 20\%).  This is primarily due to a lack of high mass
galaxies ($M_{\rm H} > 10^{11}$ M$_\odot$) which should dominate reionization at
late times within this small volume \citep{Yeh2022}. We remind readers that the
escape fraction for stars was kept fixed at the value of $f_{\rm esc} = 0.37$ from
the main, $(95.5~{\rm cMpc})^3$, {\textsc{thesan}} volume, which was calibrated to match
the observed reionization history in that more representative box. As such, we do not
expect that these small volumes necessarily match the observed global reionization trends.

All of the {\textsc{thesan}} models stand in stark contrast to the Uniform UVB
models (the two volumes overlap on this figure) which shows an instantaneous
drop in the {\textsc{Hi}} fraction to almost zero at $z_{\rm reion} = 10$, as expected. The
only {\textsc{Hi}} gas that can remain at $z < 10$ is gas that is self-shielded, making up
a very small fraction of the total gas mass in the volume (and hence an even smaller
fraction by volume).

We further demonstrate the evolution of key macroscopic quantities in the
multi-panel Fig. \ref{fig:combinedboxes}, which paints an overall picture of a
very different IGM between our two reionization models. From top to bottom, we
show the mass-weighted gas metal mass fraction, gas projected density,
mass-weighted HI fraction, and mass-weighted gas temperature projected through
the entire L4N512 volume. Different columns demonstrate these maps at different
epochs, and within each panel the top left shows the {\textsc{thesan}} volume,
and the bottom right the `mirror image', but for the Uniform UVB model.

\begin{figure*}
    \centering
    \includegraphics{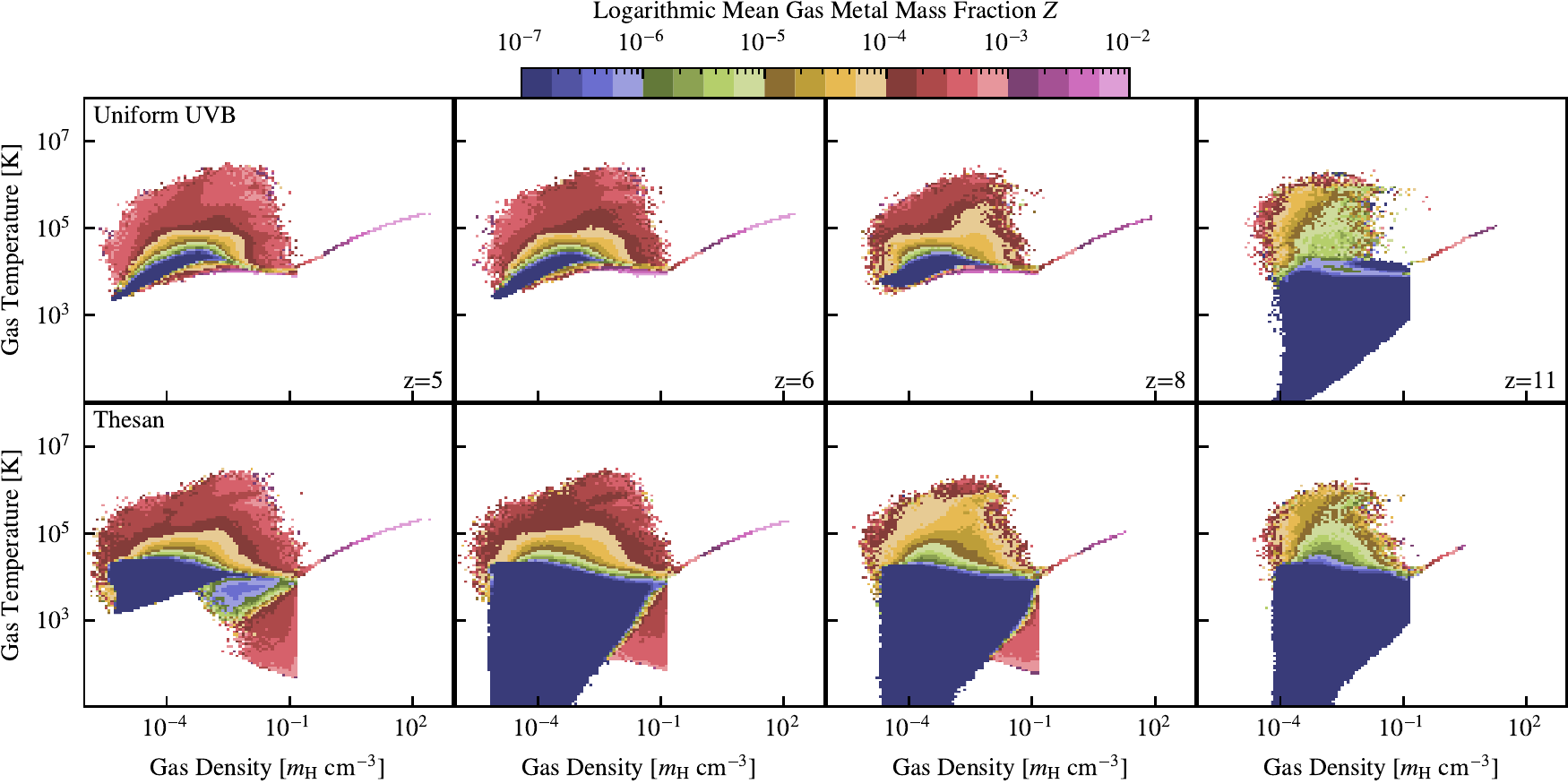}
    \caption{Phase-space diagrams, showing the temperature and physical density
    distribution of gas in the L4N512 simulations (rows), at various redshifts
    (columns). Pixels are coloured by the mean metal mass fraction ($Z$) in the
    bin, which is averaged logarithmically (i.e. $Z = \exp{\rm mean}_i \log
    Z_i$, where $i$ is the particle index), with white pixels containing zero
    gas cells.  Notable is the presence of cold, low density, gas at redshifts
    $z> 10$ for both models, which is removed instantaneously at $z=10$ by the
    reionization model in the Uniform UVB simulation. This cold, neutral, gas
    remains in the {\textsc{thesan}} simulation until the entire volume
    reionizes (approximately $z=5.6$).}
    \label{fig:rhot}
\end{figure*}

The Uniform UVB model takes an early lead in polluting the IGM with metals (at
$z=11$) thanks to increased star formation rates relative to {\textsc{thesan}},
but {\textsc{thesan}} catches up with both maps matching well at $z=5$, albeit
with more metal pollution in voids in {\textsc{thesan}}. In the second row, we
see the main cause for this increased void region pollution; the slower (and
weaker) reionization process in {\textsc{thesan}} allows dense structures to
occur at much smaller spatial scales than in a universe with a Uniform UVB, with
filaments much more prominent in the underdense regions that are all but washed
out by photoheating from the UVB.

The slow roll of reionization is abundantly clear in the third row of figures
showing the {\textsc{Hi}} fraction as a function of time. In {\textsc{thesan}}, the void in
the top of the panel does not fully reionize until $z\approx 5.5$, leaving a
vast majority of the volume neutral until this time. In addition, we see a large
abundance of self-shielded regions remain even at $z=5$ in this underdense
region, as well as there being strong self-shielding amongst the filamentary
structure. By contrast, the instantaneous reionization in the Uniform UVB model
shows a much simpler history, with only a small fraction of self-shielded
regions able to form at $z\approx 5$, with there being far fewer self-shielded
filaments than in {\textsc{thesan}} at this time, and a complete lack of small
self-shielded haloes.

The final panel gives the temperature evolution of the universe, which shows the
impact of early photoheating from the first galaxies in {\textsc{thesan}}. The
panel showing $z=11$, before reionization in the Uniform UVB universe, shows
a more extended hot circumgalactic medium (CGM) around {\textsc{thesan}} galaxies
due to additional radiative feedback \citep[see Fig. 4 in][]{Garaldi2022}.  At
$z=9$, however, this trend is reversed as the mean intergalactic medium (IGM)
temperature is pushed above $T_0 > 10^4$ K by the strong ionizing radiation
field in the Uniform UVB, whilst in {\textsc{thesan}} it can remain below $10^4$
K until $z < 7$, even in the full {\textsc{thesan-1}} volume with orders of
magnitude more massive (and hence bright) galaxies \citep{Kannan2022}.

In Fig. \ref{fig:rhot} we consider the phase space evolution of the two L4N512
simulations. The top row shows the relationship between gas density and
temperature, coloured by the metal mass fraction of the gas, for the Uniform
UVB simulation. The bottom row shows the same, but for the {\textsc{thesan}}
simulation.
 
The early lead in metal pollution that we saw in the top panels of Fig.
\ref{fig:combinedboxes} is confirmed here, with low-density ($n_{\rm H} \approx
10^{-4}$ cm$^{-3}$) hot ($T > 10^4$ K) gas having a higher metal mass fraction at
$z=11$.

The signature of the instantaneous reionization of the Uniform UVB model is
again clear, with primordial gas heated to $T \gtrsim 10^4$ K at $z=10$, a
process that is not fully repeated in {\textsc{thesan}} even at $z=5$.
{\textsc{thesan}} has a population of intermediate metallicity $Z \approx
10^{-4}$ gas at $T < 10^3$ K and high density (i.e. self-shielded {\textsc{Hi}} gas) even
down to $z=5$, a component that does not exist in the Uniform UVB simulation.

In both cases the ISM ($n_{\rm H} > 10^{-1}$ cm$^{-3}$) evolves similarly, though
it extends to higher densities early on in the Uniform UVB model due to the lack
of photoheating feedback from early stars. This is expected as we use the
equilibrium model from \citet{Springel2003} for our ISM for consistency with
{\textsc{thesan}} and other large-volume cosmological simulation suites like
EAGLE and IllustrisTNG \citep{Schaye2015, Pillepich2018}.

In summary, these first visualisations show how the IGM, CGM, and ISM evolve
differently on a global scale within the two reionization models.
{\textsc{thesan}} allows colder, self-shielded, and metal-poor gas to exist to
significantly later times than the Uniform UVB model, and this will undoubtedly
have a back-reaction on galaxy evolution.
\section{Impact on galaxy properties}
\label{sec:galaxies}

The chosen reionization model only interacts directly with the gas phase,
but this may then back-react on both the stellar phase and dark sector.
In this section we explore the impacts of the chosen reionization model
on the properties of galaxies within the simulation, in an effort to
further understand this back-reaction on the whole population.

\begin{figure}
    \includegraphics{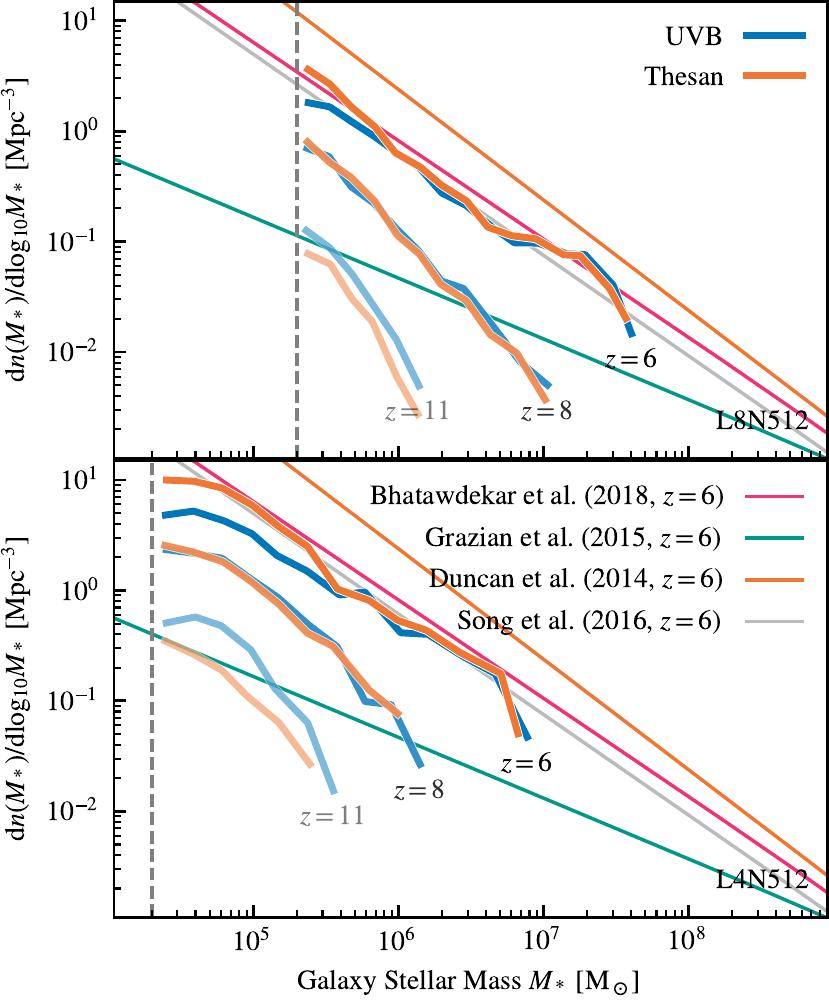}
    \caption{Galaxy stellar mass function in both simulation boxes (L8N512 in
    the top panel, L4N512 in the bottom panel), for both models (Uniform UVB
    shown in blue, and {\textsc{thesan}} shown in orange). {\textsc{thesan}}
    shows a much higher abundance ($\sim 0.5$ dex) of low mass ($M_* < 10^{6.5}$
    M$_\odot$) galaxies. The vertical grey line shows the mass of a single star
    particle in both models. Various line transparencies show the evolution with
    redshift, with the corresponding redshifts noted next to the lines at the
    same level of transparency. Vertical offsets of 0, $-0.5$, and $-1$ dex are
    applied at redshifts $z=6$, 8, and 11 respectively for clarity. In the
    background, we show fits to data from \citet{Duncan2014,
    Grazian2015,Song2016}; and \citet{Bhatawdekar2019}.}
    \label{fig:smf}
\end{figure}

In Fig. \ref{fig:smf}, we show the stellar mass function of galaxies over time
in both simulations. The top panel gives the mass function in the L8N512 volume,
with the bottom panel showing the same but for the L4N512 volume. In both cases
we see that low-mass substructures ($M_* < 10^6$ M$_\odot$) are significantly
suppressed in the Uniform UVB model, with there being as much as a $\sim 0.5$
dex difference relative to {\textsc{thesan}}. This difference persists even to low redshift,
as these low mass galaxies typically have formed early (and cannot form at later
times, as the {\textsc{thesan}} volume fully reionizes).

At the highest masses, $M_* > 10^7$ M$_\odot$, both models predict a similar
abundance of galaxies, with even small-scale box-dependent structure in the
SMF being preserved (e.g. the kink at $M_* \approx 10^{6.5}$ M$_\odot$ in the
L8N512 volume at $z=6$), indicating that high-mass galaxies may evolve similarly
between the two simulations. This is expected, as these galaxies have grown mainly
in an era where they have reionized themselves (inside-out reionization), giving
results similar to the fully ionized case. 

For comparison purposes, we show extrapolated Schechter fits from four
observational data collections, collated in \citet{Bhatawdekar2019}. All fits
here are extreme extrapolations, as the lowest mass galaxies traced by these
observations in the Hubble Frontier Fields, Hubble Ultra Deep Field, and CANDELS
surveys are around $M_* \sim 10^7$ M$_\odot$ (thanks to gravitational lensing of
background galaxies), roughly corresponding to the \emph{highest} mass galaxies
in the model. We generally show a good match to the most recent observations
from \citet{Song2016} and \citet{Bhatawdekar2019}, matching with the fits well
down to $M_* \sim 10^5$ M$_\odot$ for the {\textsc{thesan}} models (with there being an
under-abundance of galaxies of this mass in the UVB models).

Earlier data from \citet{Duncan2014} significantly overshoots our models, though
this data uses just the CANDELS South field \citep{Guo2013} and is complete
only down to $M_* \sim 10^9$ M$_\odot$, leaving a lot of freedom in the low-mass end
of the fit (with a relative 1$\sigma$ error of 50\% in the low-end slope $\alpha$,
where newer observations have constrained this to within $10\%$). Additional
early data from \citet{Grazian2015} significantly undershoots our models, but again
was complete only down to $M_* \sim 10^9$ M$_\odot$, and provides a low-end slope
that is in tension with the more recent observations of \citet{Song2016} and
\citet{Bhatawdekar2019}.

Given that the TNG model parameters were tuned to fit low-redshift ($z\sim 0.1$)
data \citep{Pillepich2018}, this level of agreement at high redshift and high
resolution when the full {\textsc{thesan}} reionization model is included is
remarkable. Future observations (and associated modelling) from upcoming
\emph{JWST} surveys, including \emph{JADES}, will allow for even tighter
constraints on the abundances of galaxies at high redshift.

\begin{figure*}
    \includegraphics{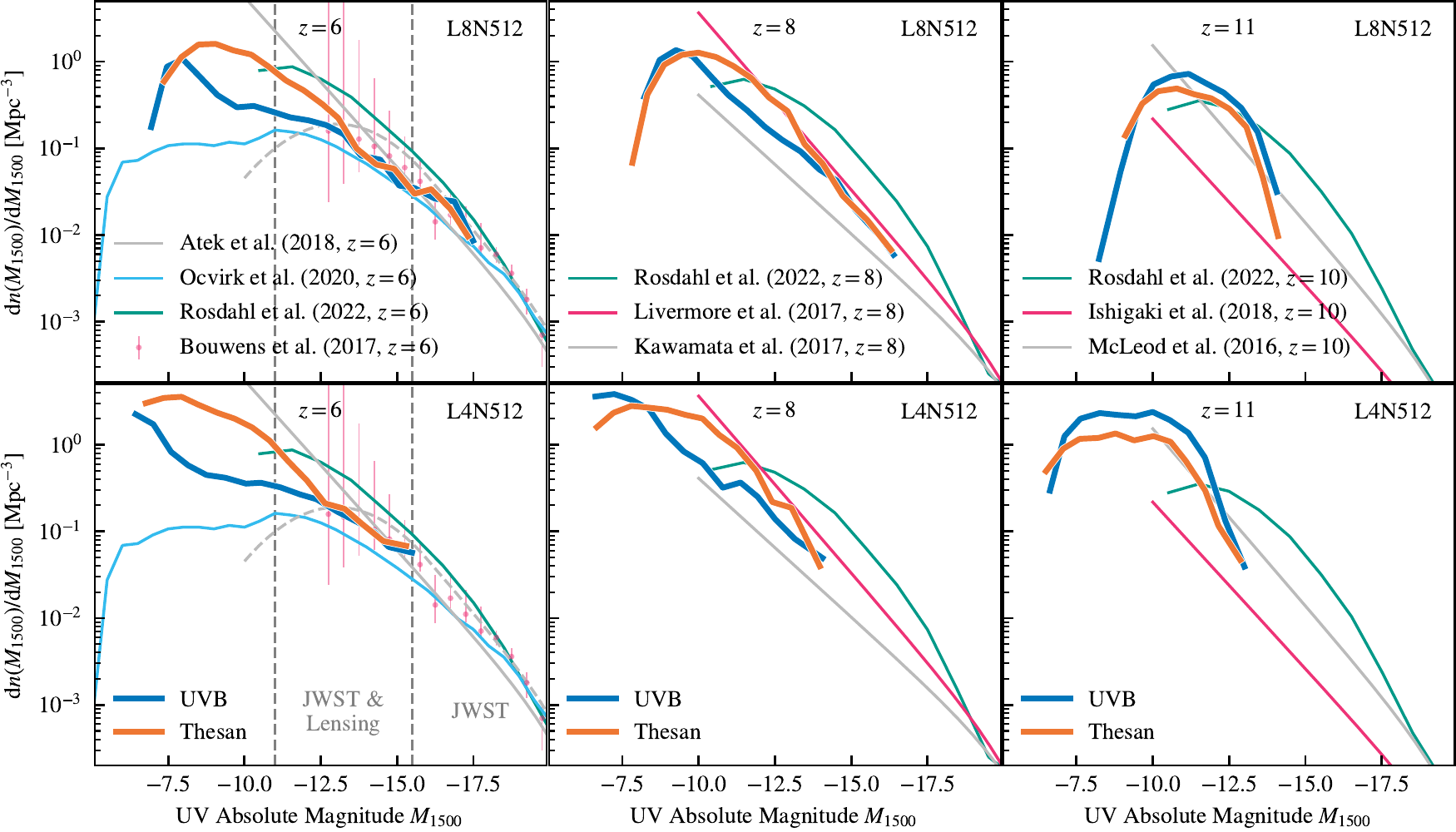}
    \caption{The UV luminosity function (in the rest-frame $M_{1500}$ band) for
    the two boxes (with L8N512 in the top row and L4N512 in the bottom row), and
    the two models (Uniform UVB shown in blue, and {\textsc{thesan}} shown in
    orange). Each column shows a different redshift, with the corresponding
    redshifts at the top of each column.  There is a significant under-abundance
    of galaxies with $M_{1500} < -14$ in the Uniform UVB model which appears to
    be converged with resolution. In the background, we show comparison
    observational data from \citet{Bouwens2017} and \citet{Atek2018} (dashed
    line shows a fit that allows for a low-brightness turn-off, solid includes
    no such component), and comparison simulation data from \citet{Ocvirk2020}
    and \citet{Rosdahl2022} at $z=6$. At $z=8$ we again show the SPHINX
    simulation in green, with fits from \citet{Livermore2017} and
    \citet{Kawamata2018} in pink and grey respectively. Finally, at $z=11$, we
    show comparison data at $z=10$ (we choose to show $z=11$ from the simulation
    to ensure a pre-reionization state in the UVB model) from SPHINX, and fits
    to data from \citet{Ishigaki2018} and \citet{McLeod2016} in pink and grey
    respectively.} \label{fig:uvlf}
\end{figure*}

In Fig. \ref{fig:uvlf} we show the UV luminosity function (UVLF), which shows the
abundance of galaxies as a function of their UV luminosity at $1500\,\si{\angstrom}$
(here denoted as $M_{1500}$). These magnitudes are generated as in
\citet{Smith2022}, using the Binary Population
and Spectral Synthesis code \citep[BPASS version 2.2.1][]{Eldridge2017}. A correction
is then applied for galaxies with poor sampling of their star formation histories
\citep[see Section 3 of][]{Smith2022}.

The figure here is laid out similarly to the stellar mass function in Fig.
\ref{fig:smf}, with the top panel showing the UVLF for the L8N512 volume, and
the lower panel for the L4N512. We seen an even larger discrepancy between the
volumes simulated with the full {\textsc{thesan}} model and the Uniform UVB
models, with there being a factor $\sim 1$ dex difference in abundance of
galaxies with $M_{1500} \approx -8$. These extremely low brightness galaxies are
unlikely to be immediately observable, even with \emph{JWST}, though there is
some hope that their abundances may be understood in lensing fields.  Following
\citet{Jaacks2019}, we show  we show the approximate absolute UV magnitude
limits for \emph{JWST}, assuming a limiting $M_{\rm 1500}$ magnitude of $M_{\rm
1500} \approx 32$, which is $M_{\rm 1500} \approx -11$ at $z=6$. 

Similarly to the stellar mass functions, the brightest galaxies have almost
equal abundances between the two models, likely due to the fact that they
reionized internally. We show excellent agreement with observational data from 
\citet{Bouwens2017} at $z=6$, and the extrapolated no-turnover model from 
\citet{Atek2018}. There is corresponding agreement with the well-known
observations from \citet{Livermore2017} and \citet{Ishigaki2018}, though these
are omitted from these figures for clarity at $z=6$. Our {\textsc{thesan}}
models exhibit significantly less of a turnover at $M_{1500} \approx -13$, as
expected from HST lensing fields, with our Uniform UVB models showing some level
of turnover. This suggests that there may be significant differences between
fields at these low magnitudes simply due to differences in reionization
history. Fields that reionized quickly as they host a massive galaxy are likely
to prefer a UVLF with a turnover (i.e. galaxy formation in very low mass haloes
is suppressed), but those in isolated regions hosting only low mass galaxies
that remain shielded will have high abundances of low-mass, and hence faint
mini-haloes and protogalaxies. The {\textsc{thesan}} and UVB models clearly
show significantly larger discrepancies in their UVLFs than their SMFs, a
consequence of their different star formation histories. These differences
will be explored further in Section \ref{sec:gas}.

\begin{figure*}
    \includegraphics{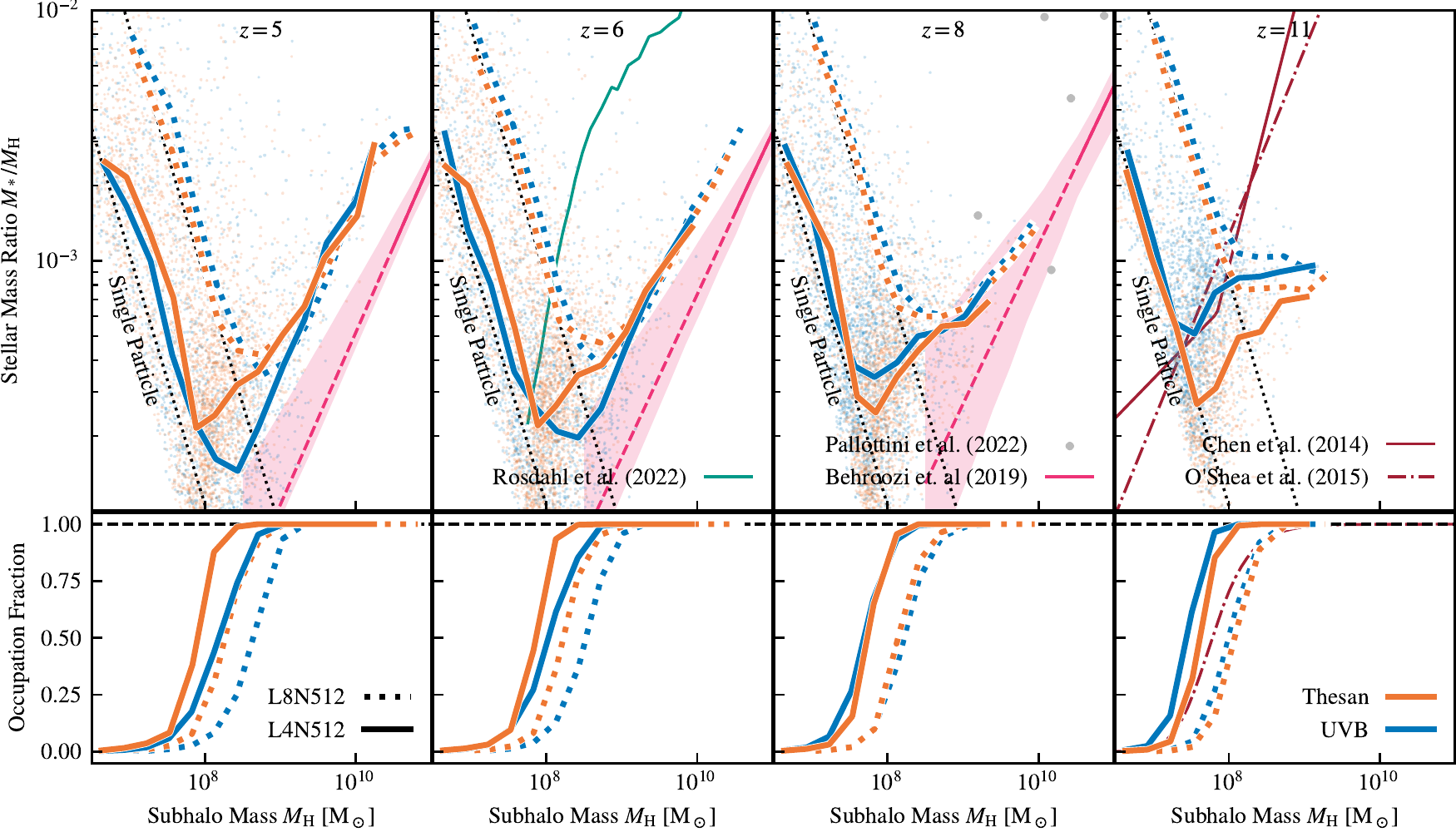}
    \caption{The evolution of the stellar mass-halo mass ratio, across cosmic
    time (different panels, top row) for each model (differently coloured lines
    for {\textsc{thesan}} and UVB, different line styles for different volumes,
    refer to the caption). The black dotted lines show the stellar mass ratio
    for a single stellar particle in the given halo for the two models (L8N512
    on top, L4N512 for the lower dotted line).  In the background, in the
    corresponding colours, we highlight the scatter in these relationships for
    the L4N512 volume.  All models are convergent at high masses ($M_H >
    10^{10}$ M$_\odot$), but show significant differences at low masses, with
    notably the L4N512 volume showing a $\sim 0.3$ dex difference in $M_* / M_H$
    at $M_H \approx 10^9$ M$_\odot$.   We also show results from
    \citet{Rosdahl2022} (green line, $z=6$), \citet{Pallottini2022} (grey
    points, $z=8$), \citet{Chen2014} (red solid line, $z=11$),
    \citet{O'Shea2015} (red dot-dash line, $z=11$), and \citet{Behroozi2019}
    (pink line and shaded region; dashed line shows where fits are extrapolated
    due to minimal mass of $10^{10.5}$ M$_\odot$ in this work). The lower panels
    highlight differences in occupation fraction, defined as the fraction of
    haloes within the bin that host at least a single star particle, as a
    function of halo mass. Though the occupation fraction does not display
    convergence with simulation resolution, we do see that the {\textsc{thesan}}
    model predicts higher occupation fractions at lower halo masses than the
    Uniform UVB model. }
    \label{fig:smhm}
\end{figure*}

We also show comparisons to two recent simulations: CoDa~II \citep{Ocvirk2020},
and SPHINX-20 \citep{Rosdahl2022}. SPHINX provides a comparable luminosity
function to {\textsc{thesan}}, with it showing no significant turnover at low
luminosity\footnote{The increased abundance of all galaxies at low masses is
likely due to higher stellar masses across the board for SPHINX; see the
discussion around Fig. \ref{fig:smhm}.}. The results from CoDa II do show
significant turnover at $M_{\rm 1500} \approx -11$, with this suppression being
dependent (as in {\textsc{thesan}}) on the ionizing radiation background. At
these low magnitudes, CoDa II struggles to resolve haloes, with its minimal
resolved halo mass $M_H \approx 10^7$ M$_\odot$ having an expected UV luminosity
$M_{\rm 1500} \approx -8$, meaning this turnover is potentially numerical in
origin. CoDa II, due to its fixed grid strategy, also struggles to resolve
processes within galaxies, with physical resolution at $z=6$ larger than
$3$ kpc. Our L8N512 model, which has similar mass resolution, but better spatial
gas resolution due to our adaptive technique, also turns over in all cases
around this magnitude, a non-converged effect that goes away as we move to the
higher resolution L4N512 volume.

Additionally, the results presented here can be compared with those from the
Cosmic Reionization on Computers (CROC) project, specifically the predictions
for the faint-end of the UVLF presented in \citet{Gnedin2016b}. Both simulation
suites, despite using vastly different numerical methods, provide good fits to
the available UVLF data. Notably, we both find that the turn-over in the UVLF at
faint magnitudes is not reliably predicted by our models, with it being
dependent on both the reionization history of the volume (the difference between
our orange and blue lines), as well as the specific details of the star
formation modeling within the simulations (highlighted here by different
turn-over positions at the two different resolution levels).

We explore these trends further in Fig. \ref{fig:smhm}, which shows both the
stellar-to-halo mass ratio and subhalo stellar occupation fraction trends with
mass for all of our models. We choose to show the total stellar mass bound to
the subhalo $M_*$ and total halo mass bound to the subhalo $M_{\rm H}$ here, for
more straightforward comparisons \citep{Rosdahl2022}, a departure from the main
{\textsc{thesan}} papers. All subhaloes are shown in the background as
appropriately coloured scatter.

Similarly to our abundances, we see that the stellar-to-halo mass ratio
$M_*/M_{\rm H}$ is well converged between simulation resolutions and even
reionization models at all redshifts for masses $M_{\rm H} \gtrapprox
10^9$~M$_\odot$. As in \citet{Kannan2022} our models match well with the
abundance matching results from UNIVERSEMACHINE \citep{Behroozi2019} at all
valid redshifts ($z < 10$). Our haloes are generally in the regime where the
UNIVERSEMACHINE fits must be extrapolated, as their fits are only valid down to
$M_{\rm H} > 10^{10.5}$ M$_\odot$ (indicated by the dashed part of the pink line
in the figure). Our galaxies also match well with the zoom simulations of
\citet{Pallottini2022} at $z=8$, though their lowest mass haloes typically
correspond to the highest masses that are available even in our lower resolution
L8N512 volume. We find, consistent with the CROC simulation suite, an increasing
stellar-to-halo mass relation with time \citep{Zhu2020}. This is consistent with
the simple model results from \citet{Tacchella2018}, which relied on an assumption
that star formation is delayed relative to gas in-fall.

At $z=11$ we show the fit from \citet{Chen2014} (at $z=15$ from the Rarepeak
simulations), which was shown to match well with extrapolated data from
\citet{Behroozi2013}. At these very early epochs, the data from \citet{Chen2014}
does match well with our background scatter, but we cannot achieve the high
levels (1\%) of star formation efficiency at halo masses of $M_{\rm H} \approx
10^9$ M$_\odot$ that they find. Using the same model and at the same redshift of
$z=15$, we show the results from \citet{O'Shea2015}, which again shows very high
star formation efficiencies, meaning that differences here are likely both
resolution and model-dependent.

Finally, at $z=6$, we compare to the results from the SPHINX-20 volume
\citep{Rosdahl2022}.  We find stellar masses roughly an order of magnitude lower
than theirs that are in tension with the aforementioned observational and prior
simulation work due to significantly higher star formation efficiencies. Such differences
are expected given that the SPHINX model was designed to simulate galaxies at redshifts
$z>5$, and when it is used to trace galaxies to lower redshifts ($z\approx3$) it
has been shown to produce bulge-dominated galaxies \citet{Mitchell2021}, whereas
THESAN uses the IllustrisTNG model which was calibrated against the stellar-to-halo
mass ratio at $z=0$ \citep{Pillepich2018}.

Our results for the UVLF and SMF can also be compared to \citet{Wu2019} who also
employed {\textsc{Arepo-RT}}, though with the Illustris galaxy formation model
\citep{Vogelsberger2014}, a predecessor to our IllustrisTNG model. Our galaxy
stellar mass functions appear broadly similar, and our high-luminosity UVLFs
both show excellent agreement with the \citet{Bouwens2017} data at $z=6$.
\citet{Wu2019} did not simulate to a high enough resolution to model galaxies
with $M_* < 10^6$ M$_\odot$ (cutting off their UVLF at $M_{1500} = -15$) and as
such found that the UVB and fully self-consistent reionization models were
similar on these metrics, whereas we have expanded the parameter space to
include lower mass galaxies in our analysis.

Differences between reionization models are largest in the highest resolution
volume, which is able to well resolve star formation in haloes with mass
$10^{7.5} < M_{\rm H} / {\rm M}_\odot < 10^9$. The {\textsc{thesan}} models have
significantly (a factor $\sim 2$ at $z \approx 5$) higher stellar-to-halo mass
ratios in these haloes, indicating a higher star formation efficiency. This is
also seen to a much reduced extent in the L8N512 volume, where modelling the
star formation histories is limited by poor sampling (the dotted black lines
indicate when a single stellar particle resolves the star formation history of
the entire halo). The upturn in efficiency at this mass scale is not physical,
but numerical in nature, which is clear when comparing results from 
the L8N512 and L4N512 volumes.

In the lower panels, we show the (stellar) occupation fraction of the haloes.
This is calculated as the fraction of haloes within a given $M_{\rm H}$ bin that
host at least one star particle. There is a clear resolution-dependent trend
here, indicated best at $z=11$, where the higher resolution (L4N512) simulations
can host star particles at lower halo masses than the lower resolution volumes
(L8N512). As the simulation evolves, however, we see reionization-dependent
behavior; the {\textsc{thesan}} models always host star particles at lower halo masses
than their Uniform UVB counterparts at low redshifts ($z < 6$), further
indicating that the Uniform UVB models suppress star formation in low mass haloes
($M_{\rm H} < 10^9$ M$_\odot$).
We see relatively good agreement for our two models with the occupation fractions
reported by \citet{O'Shea2015}, with both of us reporting no stars in haloes
with virial masses $M_{\rm H} \lessapprox 10^7$ M$_\odot$.

\begin{figure*}
    \includegraphics{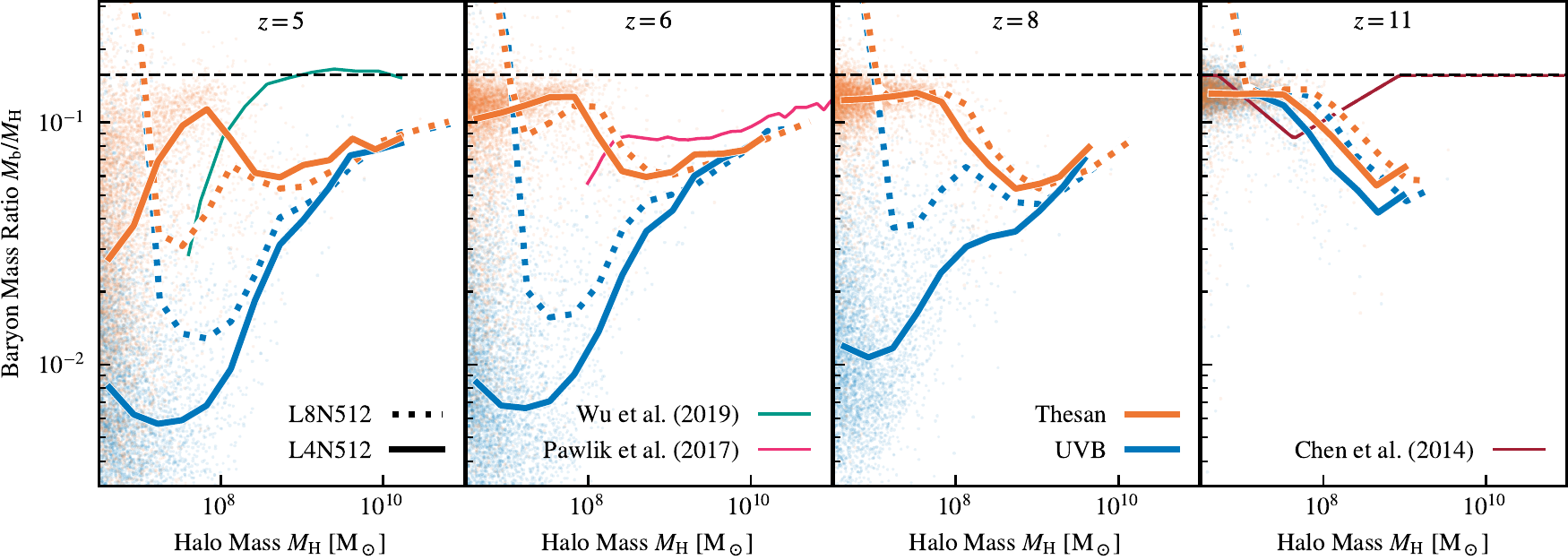}
    \caption{The evolution of the bound baryon mass-halo mass ratio, across
    cosmic time (different panels) for each model and volume size (differently
    coloured lines showing the median trend for the two reionization models,
    with different thick lines showing the different volumes). The dashed line
    shows the universal baryon fraction, corresponding to $\Omega_{\rm b} /
    \Omega_{\rm m} = 0.157$. In the background, the appropriately coloured
    points show the scatter in the relation for the L4N512 volumes (each point
    represents a single subhalo). We see that the Uniform UVB models rapidly
    lose baryons post-reionization in low mass haloes ($M_{\rm H} < 10^9$
    M$_\odot$), with the {\textsc{thesan}} volumes retaining close to the
    universal baryon fraction even down to $z \approx 5$.  In the background, we
    show results from \citet{Pawlik2017} (pink line, Aurora, L12N512, $z=6$),
    \citet{Chen2014} (red line, $z=11$), and \citet{Wu2019} (green line, $z=5$,
    Illustris model) and see good agreement between these results and the
    {\textsc{thesan}} models.}
    \label{fig:baryonmass}
\end{figure*}

To further investigate the suppression effects of the Uniform UVB model, we show
the baryon mass ratio $M_{\rm b}/ M_{\rm H} = (M_* + M_{\rm g}) / M_{\rm H}$ in
Fig. \ref{fig:baryonmass}. As this is (generally) dominated by the gas in the haloes,
with $M_{\rm b} \approx M_{\rm g} \gg M_*$, it is less sensitive to the specifics of
star formation physics (though, as seen in \citet{Chen2014}, the large scatter in this
metric is generally driven by supernova physics).

At early times, all of our models provide consistent baryon mass ratios, with
both high resolution and low resolution models hosting the universal baryon
fraction $f_{\rm b, U} = \Omega_b / \Omega_m = 0.157$ (dashed black line) at halo masses
$M_{\rm H} < 10^8$ M$_\odot$, and higher-mass haloes hosting slightly lower
baryon fractions due to long cooling times and gas pressure.

By $z=8$, we see the impact of the Uniform UVB model on the gas fractions.
{\textsc{thesan}} haloes at $M_H < 10^9$ M$_\odot$ can still retain gas to maintain $f_{\rm
b, U}$, but haloes in the Uniform UVB model have their baryons almost entirely
removed as the gas becomes unbound due to external photoheating. This halo mass corresponds
roughly to the mass at which haloes have a virial temperature,
\begin{equation}
    T_{\rm vir} = \frac{1}{2} \frac{\mu m_{\rm p}}{k_{\rm B}} \frac{GM_{\rm vir}}{r_{\rm vir}}
    \label{eqn:tvir}
\end{equation}
where $\mu$ is the molecular weight per particle for fully ionized gas, $m_{\rm
p}$ the proton mass, $k_{\rm B}$ the Boltzmann constant, $G$ is Newton's
gravitational constant, and $M_{\rm vir}$ and $r_{\rm vir}$ are the virial mass
and radius of the halo, of around $10^{4}$ K. Hence fully ionized hydrogen gas
in thermal equilibrium has enough thermal energy to escape the
gravitational potential and becomes unbound to these low mass haloes.

\begin{figure*}
    \includegraphics{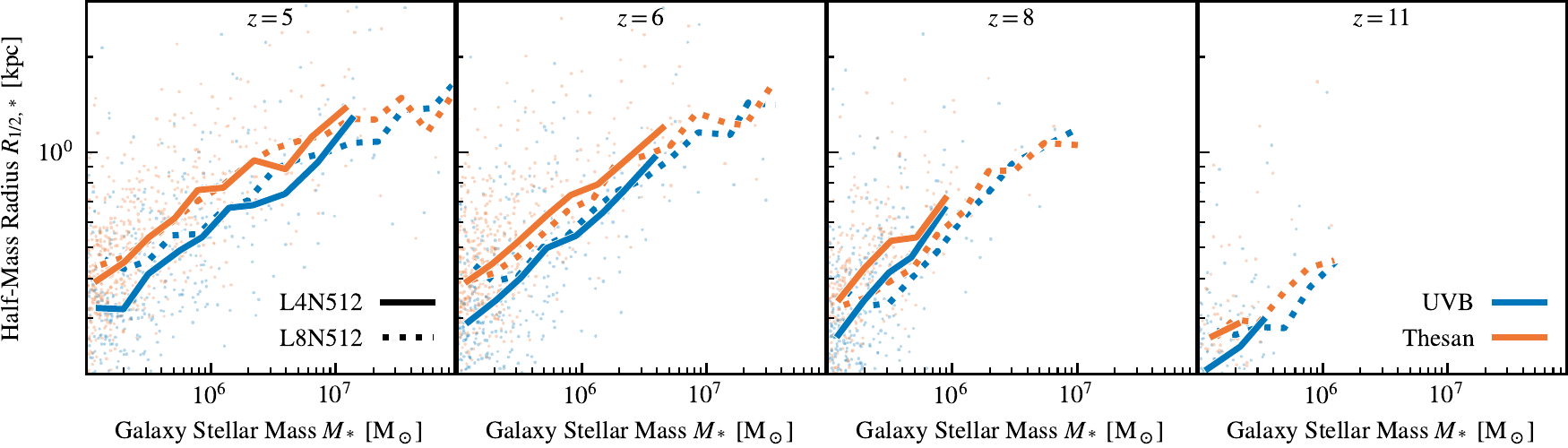}
    \caption{The half-mass stellar sizes of galaxies for both models (line
    colours) and box-sizes (line styles). Each panel shows a different redshift
    from $z=11$ to $z=5$. Typically the {\textsc{thesan}} model leads to stellar
    sizes that are roughly $\sim 2$ times as large as the Uniform UVB model,
    indicating significant differences in galaxy morphology and evolution that
    are dependent on the reionization history of both the central galaxy and its
    numerous progenitors.}
    \label{fig:sizes}
\end{figure*}

As haloes of these low masses are slowly externally reionized, the
{\textsc{thesan}} models see a reduction in the baryon fraction at low masses
all the way down to $z=5$. As previously discussed, at $z=5$ there is still a
small fraction of our L4N512 ($x_{\rm HI} \approx 0.05$) and L8N512 ($x_{\rm HI}
< 0.01$) volumes that are neutral, meaning that the baryon fractions still do
not match the Uniform UVB models at this time. There is significant down-scatter
from the universal baryon fraction in the {\textsc{thesan}} model here, with
this spread representing haloes that are in the process of being externally
reionized.

At $z=6$ we are able to compare with prior simulation results from the Aurora
simulations presented in \citet{Pawlik2017}. We match well, to within $0.1$ dex
for our medians at high masses. The Aurora simulations do not have the
resolution (their highest resolution volume would be L12N512 in our parlance) to
fully capture the transition to haloes with virial temperatures lower than
$10^4$ K as we do, but they maintain high baryon fractions comparable to our two
{\textsc{thesan}} models at this redshift.

At $z=5$ we show the results from \citet{Wu2019}, which used the Illustris
galaxy formation model. We see broadly similar trends to our own
{\textsc{thesan}} models, though the Illustris model is able to retain universal
baryon fractions at $M_{\rm H} > 10^{8.5}$ M$_\odot$, despite their use of
$f_{\rm esc} = 0.7$, indicating inefficient stellar feedback.

Fig. \ref{fig:sizes} explores another key galaxy scaling relation: the mass--size
relationship. Here we show the mass contained within a spherical aperture of
size twice the stellar half-mass radius, with the stellar half-mass radius
calculated as the radius of a spherical aperture containing half of the bound
stellar mass to each subhalo.

Galaxies at these high redshifts and low masses have small sizes, typically
around $R_{1/2, *} \approx 1$ kpc, and have not yet formed disc-like
morphologies.  We see that the differences in baryonic accretion and retention
translate into differences in the stellar morphologies, with {\textsc{thesan}} galaxies
having sizes typically $\sim 0.1$ dex larger, consistent with less rapid early
star formation and lower levels of bulge formation \citep{Crain2015}. This is 
true across all epochs post-reionization, with {\textsc{thesan}} galaxies still maintaining
larger sizes at $z=11$ due to the additional radiative feedback \citep{Wise2008,
Wise2012}.

There are no reliable observations of galaxy sizes of these masses available for
comparison, but we do see that our galaxy sizes (at mass scales where the
simulations overlap) are consistent with those found in TNG-50 \citep[although
those results are only available at $z=2$ in published work]{Pillepich2019}.

\begin{figure}
    \centering
    \includegraphics{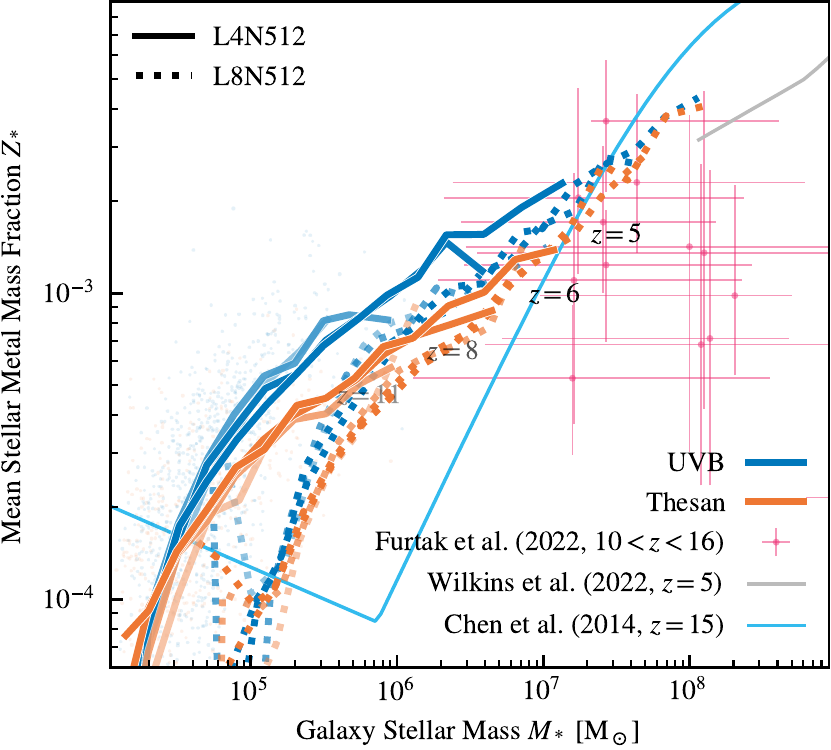}
    \caption{The mean stellar metal mass fraction of stars within galaxies ($R <
    2 R_{1/2, *}$) as a function of their mass within the same aperture. Solid
    lines (and background scatter) show the median trend in the higher resolution
    L4N512 volume, and dotted lines show the lower resolution L8N512 volume.
    Different transparency levels within lines show different redshifts, with
    labels appropriately transparent. We see that the Uniform UVB model leads to
    systematically ($\sim 0.3$ dex) higher stellar metallicities within
    galaxies, though the L4N512 volume shows a greater degree of separation.}
    \label{fig:stellarmetallicity}
\end{figure}

Finally we turn to the metallicities of our galaxies. Fig.
\ref{fig:stellarmetallicity} shows the stellar metal mass fractions ($Z_*$) of
our galaxies as a function of their stellar mass. We see here that the Uniform
UVB simulations generally form galaxies with significantly higher metallicities
than the {\textsc{thesan}} model. This can be connected back to the phase space
diagrams in Fig. \ref{fig:rhot}, where we showed that {\textsc{thesan}} allows
for cold, low metallicity (or even primordial) gas to remain until even $z=5$,
which can accrete easily into galaxies for star formation. We note that the dust
content of the galaxies is similar between the two reionization models.
Additionally, the low-mass neutral haloes in Fig. \ref{fig:bigplot} can allow
for significant ex-situ star formation for even these massive galaxies, with
these early stars being extremely metal poor.

We show, for comparison, two models that report stellar metallicities. First,
the fit from \citet{Chen2014} shows significantly lower metallicities at low
masses than we do, but matches at $M_* \approx 10^7$ M$_\odot$. \citet{Chen2014}
uses stellar particle masses that are much lower than ours (in some cases $M_*
\approx 10^3$ M$_\odot$), and additionally include a separate model for
Pop III stars. This changes their early enrichment behaviour relative to ours.
By comparison the FLARES simulations of \citet{Wilkins2022}, that target much
higher msases, and use a Uniform UVB model, match well with our simulated
model employing a Uniform UVB.

In addition to these prior models, we show early \emph{JWST} results from SED
fitting from \citet{Furtak2022}. These observations are of galaxies lensed
behind the cluster SMACS J0723.3-7327 \citep{Atek2022} using the seven
broad-band filters from NIRCam, and one band from NIRISS, with SED fits
completed using BEAGLE.  Though the error bars are large, and these observations
are from only one field, we already see that our Uniform UVB model in our
highest resolution simulation overshoots the majority of the data. Our highest
resolution {\textsc{thesan}} model, that reliably can form self-shielded
minihaloes with $T_{\rm vir} < 10^4$ K, and hence fuel galaxies with cold
primordial gas and ex-situ metal poor stars, shows a trend that looks to be more
consistent with these early observations.  Additional observations are available
from \citet{Tacchella2022}, but the galaxy stellar masses are too high ($M_* >
10^9$ M$_\odot$) for reliable comparison to our small volumes.
\section{Understanding gas phase evolution}
\label{sec:gas}

To better understand this process of primordial gas fueling, we now turn to the
gas phase and birth properties of stars. \citet{Wise2014} and \citet{Chen2014} found
that galaxies in haloes with virial temperatures $T_{\rm vir} < 10^4$ K dominate
early galaxy formation, which appears generally consistent with our prior findings.

\begin{figure*}
    \includegraphics{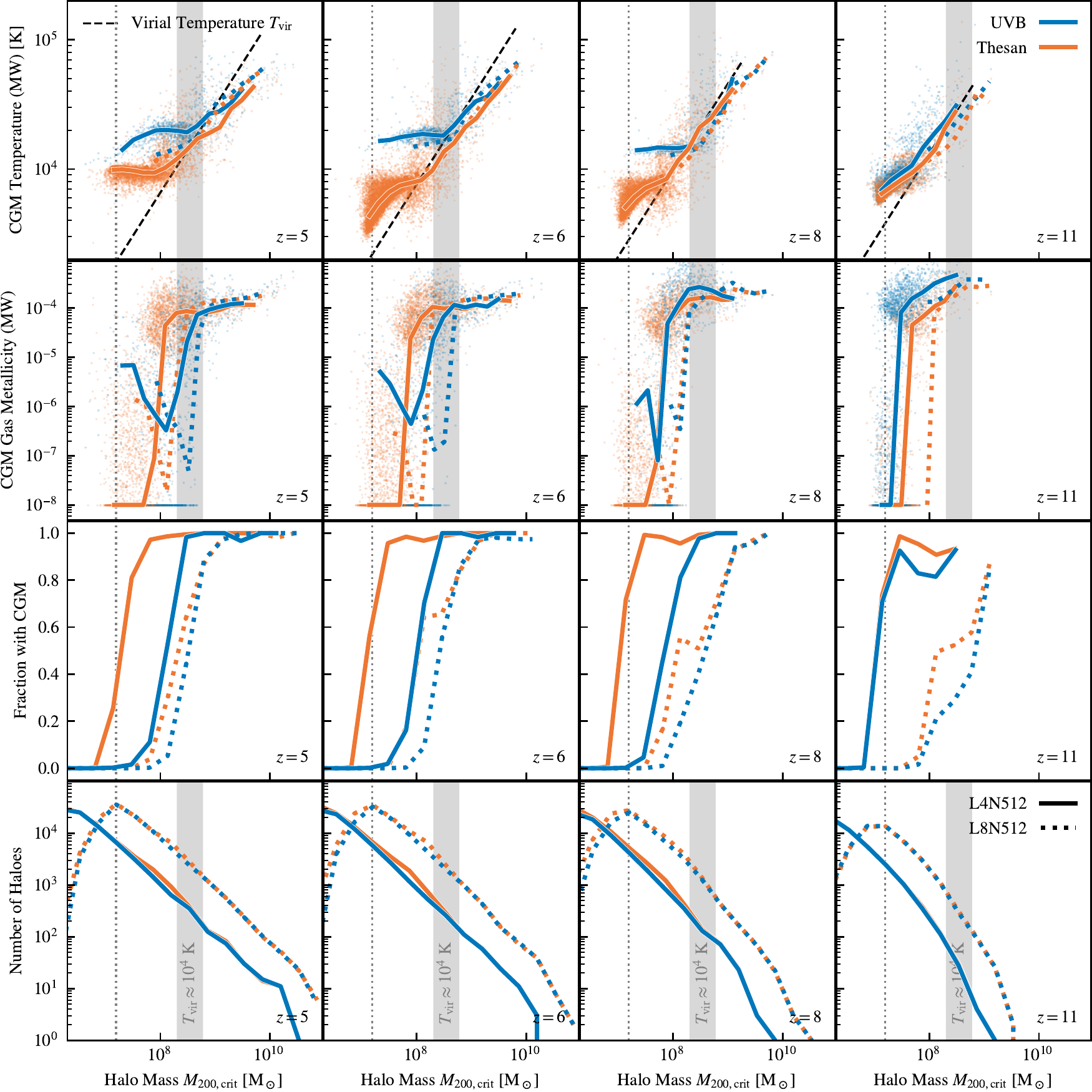}
    \caption{Trends over time of CGM properties (for central haloes only) for
    the two simulation models (Uniform UVB and {\textsc{thesan}} shown as different
    colours, with the two box sizes and resolutions shown as different line
    styles, and scatter shown for the L4N512 volume only). \emph{Top panels}:
    the CGM temperature as a function of halo mass, with the dashed line showing
    the halo virial temperature according to equation (\ref{eqn:tvir}). At high
    redshifts, all simulation models follow this line, but at halo masses
    corresponding to a virial temperature of lower than $\sim 10^{4}$ K
    systematic increases are seen with redshift as the haloes reionize.
    \emph{Second panels}: the mass-weighted metallicity of CGM gas as a function
    of halo mass. Post-reionization, haloes with masses corresponding to virial
    temperatures $T_{\rm vir} < 10^4$ K struggle to retain any primordial gas,
    as this gas is no longer bound due to its thermal energy. \emph{Third panels}:
    The fraction of haloes with at least a single bound gass particle in their CGM.
    Pre-reionization ($z>10$ in the UVB model), this fraction is  close to unity
    for all resolved haloes (denoted by the dotted line). Post-reionization,
    no halo with $M_{\rm 200, crit} \lessapprox 10^{8.5}$ M$_\odot$ can retain 
    CGM gas. \emph{Bottom panels}: The number of haloes within each mass bin, using
    the same bins as the above panels. The number of galaxies with CGM included
    is hence the bottom row multiplied by the third row.}
    \label{fig:cgm}
\end{figure*}

In Fig. \ref{fig:cgm} we show the CGM properties of all central haloes in our
simulations. To calculate the CGM properties, we take each central halo, and
only include gas that is bound to the central subhalo, with all other subhaloes
removed.  We then remove gas that is in the central `galaxy' by excluding gas
cells that lie within twice the stellar half-mass radius $R_{1/2, *}$. Filtering
spatially rather than by gas properties excludes the vast majority of ISM gas,
whilst allowing any recently accreted dense gas in the CGM to remain part of our
selection.

The top row of this multi-panel figure shows the evolution of the mass-weighted
mean CGM gas temperature with cosmic time. In the background, we show all haloes
as appropriately coloured scatter (for the L4N512 volume), and the lines
represent the median value binned as a function of halo mass.

At early times ($z=11$), the median trend closely follows the virial temperature
of the host haloes. We see significant up-scatter from the median in the Uniform
UVB, due to higher star formation rates in the highest mass haloes and hence
stellar feedback that can heat the surrounding CGM. 

Post-reionization in the Uniform UVB model we see that all haloes with $M_{\rm 200, crit} <
10^{8.5}$ M$_\odot$ have a CGM temperature consistent with the recombination
equilibrium temperature for {\textsc{Hii}} gas, as all CGM gas in this model is ionized by
the external radiation field.  The lowest mass haloes in {\textsc{thesan}}, by comparison,
retain a cold CGM with $T_{\rm CGM} < 10^4$ K (albeit with more scatter as some
photoheating occurs). This trend occurs until $z=5$ when the majority of the
region (and certainly most CGM) is reionized, where the minimal temperature of
CGM gas is $T_{\rm CGM} \approx 10^4$ K. Differences in the $z=5$ CGM temperatures
are also seen, with the UVB heating the gas to a temperature $\sim 2$ times as 
high as the {\textsc{thesan}} model. This will impact gas accretion rates and
the impact preventative feedback even now that the volume is reionized.

We see further evidence of the early star formation boost in the Uniform UVB model
through enrichment of the CGM in the second row of Fig. \ref{fig:cgm}. Here we show the
mass-weighted mean gas metal mass fraction of the CGM as a function of halo mass, after
applying a metallicity floor of $10^{-8}$. At $z=11$ the Uniform UVB model shows
stronger enrichment, with $\sim 0.5$ dex higher CGM metallicities, corresponding
to the higher stellar-to-halo mass ratio (Fig. \ref{fig:smhm}). This higher metallicity
even persists to $z=8$, but is washed out by $z=5$.

At lower masses, $M_{\rm 200, crit} < 10^8$ M$_\odot$, the CGM mainly consists
of primordial gas. This is crucial, as in strong interstellar radiation fields,
it is not possible with $T_{\rm vir} < 10^4$ K to retain gas without metal line
cooling. This is made abundantly clear in the transition between $z=11$ and
$z=8$ in the third row, which shows the fraction of haloes that have a single
gas cell bound to their CGM. At $z=11$, both {\textsc{thesan}} and the Uniform UVB
models have bound CGM down to $M_{\rm 200, crit} \approx 10^7$ M$_\odot$, but
this CGM is removed at $z=8$ for the Uniform UVB at masses $M_{\rm 200, crit} <
10^8$ K as photoheating causes gas to become unbound. This is despite there
being a large number of haloes within each bin here (over $10^3$), showing the
spatially universal power of a Uniform UVB.

This story could potentially be changed by significant pre-enrichment from early
Pop III stars in molecular cooling haloes not yet experiencing strong
Lyman-Werner radiation that prevents the formation of molecular hydrogen, which
would allow metal cooling in these low mass haloes to potentially overcome even
the Uniform UVB \citep{Wise2019}. This enrichment would need to be significant,
however, as at metallicities well below solar $Z \ll Z_\odot$, primordial cooling
typically dominates around $T\approx 10^{4}$ K \citep{Ploeckinger2020}. We can
see the impact of slow enrichment on these low mass galaxies through the
background scatter up from primordial gas in the metallicity of gas in these
haloes of $M_{\rm 200, crit} < 10^8$ M$_\odot$ from $z=8$ to $z=5$.

\begin{figure}
    \centering
    \includegraphics{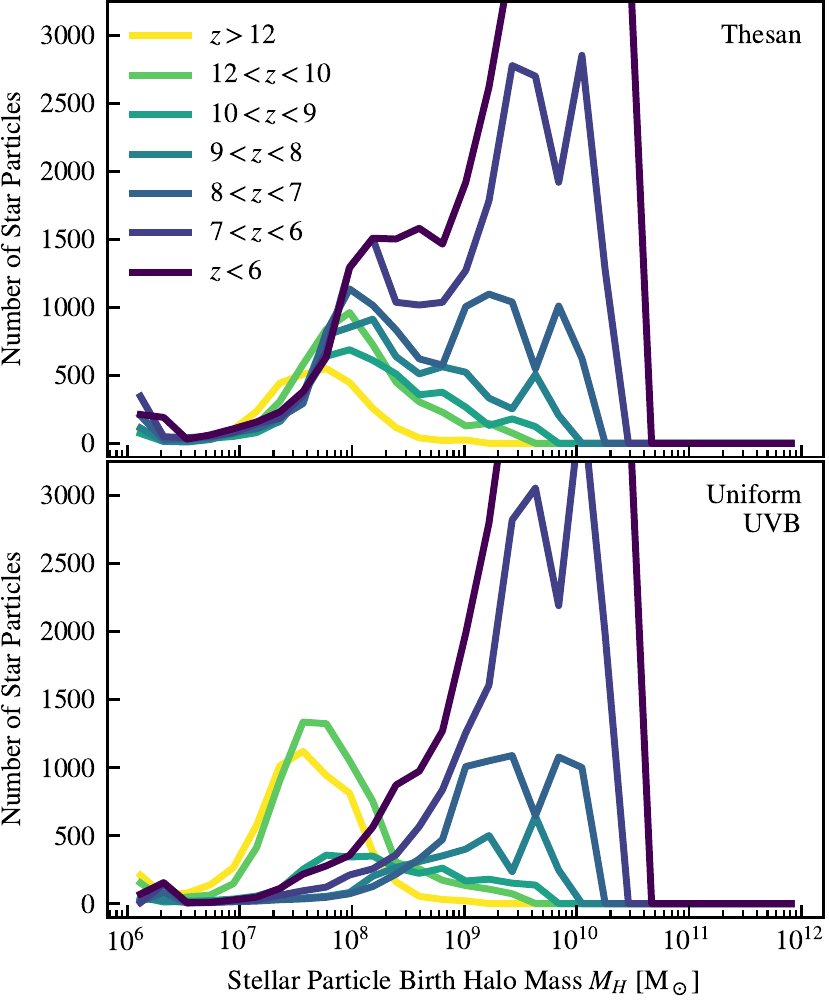}
    \caption{The distribution of birth halo masses, as a function of redshift
    (differently coloured lines) for the two models in the L4N512 box, with the
    top panel showing {\textsc{thesan}}, and the bottom showing the Uniform UVB
    model.  {\textsc{thesan}} can continue to form stars in low mass ($M_H <
    10^{8.5}$ M$_\odot$) haloes even down to $z < 6$.}
    \label{fig:stellarbirth}
\end{figure}

In Fig. \ref{fig:stellarbirth} we show how these different gas phase trends with
halo mass can impact the birth of stars in galaxies. For each stellar particle,
representing a whole stellar population, we compute both the time of its birth
(and hence redshift), as well as the halo mass within which it was first found.
This allows us to view where and when stars were born over cosmic time for
both simulations. Fig \ref{fig:stellarbirth} shows the number of stars born
in each halo, with each line representing a different redshift.

First considering the bottom panel, we see that in the Uniform UVB simulation
many stars are born in haloes with $10^7 < M_{\rm H} / {\rm M}_\odot < 10^9$ at
redshifts $z>10$. This is shut down immediately, as predicted by the CGM
properties in Fig. \ref{fig:cgm}, at the time of instantaneous reionization of
$z=10$. Comparing at these redshifts with {\textsc{thesan}}, we see that far more stars are
born in these haloes at high redshifts (a factor of $\sim 2$ more, consistent
with Fig. \ref{fig:reionizationhistory}).

By contrast, the {\textsc{thesan}} model allows star formation to continue in
these low mass haloes all the way down to $z < 6$, though at reduced levels as
parts of the volume become fully ionized. The additional star formation in these
low mass haloes is offset by there being less star formation (i.e. the number of
stars born in each halo, as the number of haloes of a given mass is consistent
between reionization models) in high mass haloes due to lower amounts of gas
accretion and additoinal photoionizing feedback. These two work together to
match the bright end of the UVLF, SMF, and global star formation rate at $z
\approx 5$ between both {\textsc{thesan}} and the Uniform UVB model.

\begin{figure}
    \includegraphics{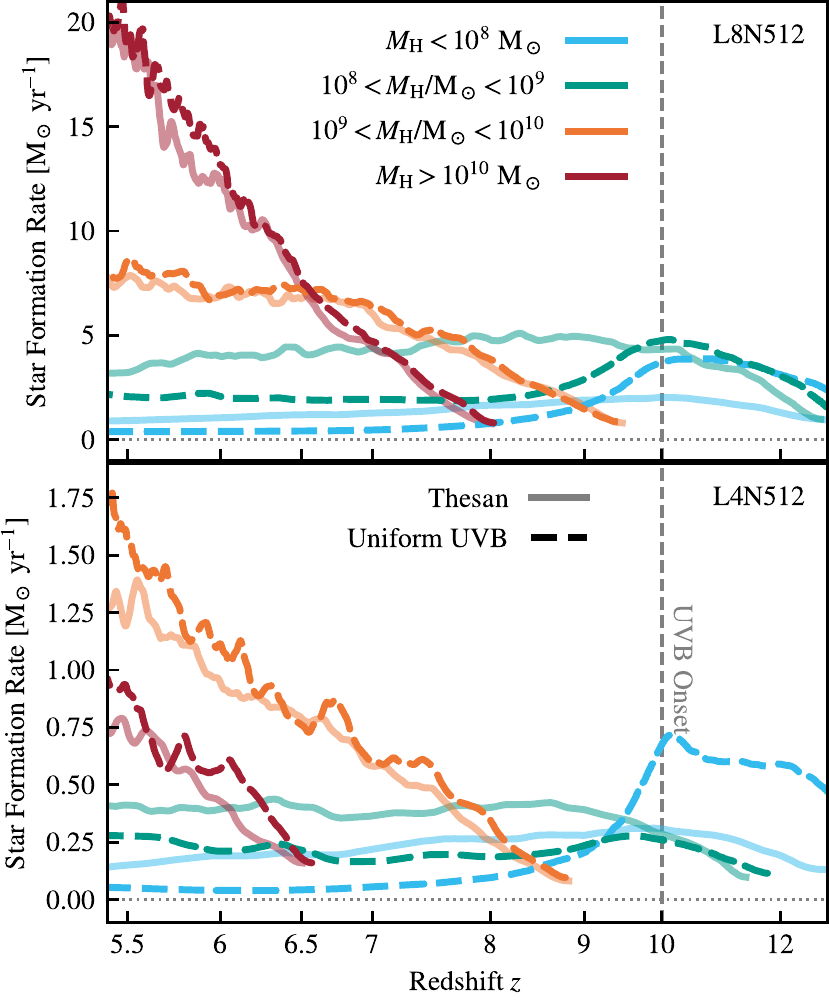}
    \caption{Star formation history, split by the birth halo mass of the stars,
    for the two box-sizes (top and bottom) and two models (line styles). The 
    vertical dashed line indicates $z=10$, when the instantaneous Uniform UVB
    begins. This star formation history was created by using the birth times of
    the stellar particles, rather than gas star formation rates, and each point
    is sampled using the birth times of the nearest (in time) 1024 star
    particles.}
    \label{fig:splitsfh}
\end{figure}

Fig. \ref{fig:splitsfh} rephrases the information in Fig. \ref{fig:stellarbirth}
to show the star formation history of the entire volume, but split by birth halo mass.
We show the L8N512 volume in the top panel, and the L4N512 volume in the lower panel.
The {\textsc{thesan}} volumes are shown as solid lines, with the Uniform UVB models shown as dashed
lines. The highest mass haloes, with $M_{\rm H} > 10^9$ M$_\odot$ have \emph{in-situ}
star formation rates that are insensitive to the choice of reioniozation model. These
galaxies will have strong enough photoionization fields originating from their own
star particles that they effectively live in completely ionized bubbles (aside from
self-shielded gas) of their own making, meaning that we do not expect differences between
models. However, should a galaxy of this mass have a significant portion of its stars
formed ex-situ in lower mass haloes, it will have an altered star formation history.

\begin{figure}
    \includegraphics{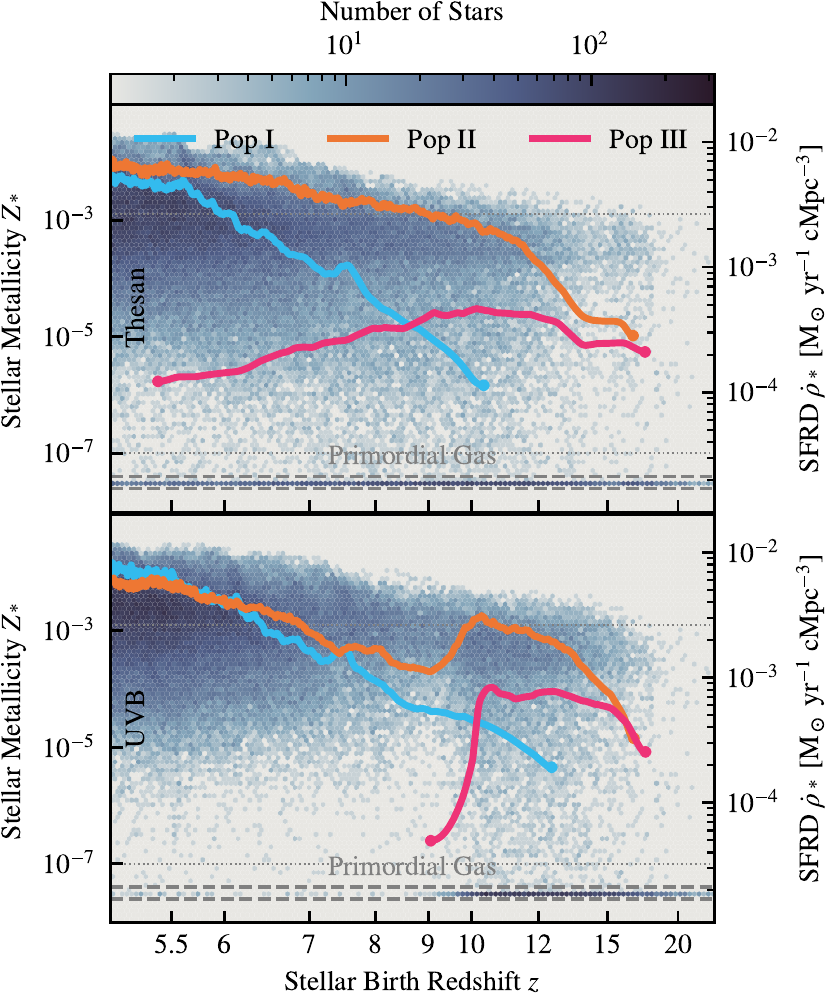} \caption{The
    distribution of stellar metallicities as a function of the birth redshift of
    the stars (background hexbins, coloured by the number of stars in each bin).
    The dashed grey lines show purely primordial gas which is given a minimal
    metal mass fraction $Z_* = 10^{-7.5}$. The secondary axis shows the global
    star formation rate density $\dot{\rho}_*$, split by stellar epoch
    (differently coloured lines; dotted grey lines show the metallicity
    thresholds). The two panels show the {\textsc{thesan}} and UVB models (from top to
    bottom), which both have very different metallicity distributions (Uniform
    UVB shows almost no low-metallicity Pop III stars below $z=10$), leading to
    significant (over $\sim 2$ dex in Pop II) differences in star formation
    rates.}
    \label{fig:metallicity}
\end{figure}

For the lower mass haloes, $M_{\rm H} < 10^9$ M$_\odot$, the Uniform UVB model 
here shows again its higher star formation rate at $z > 10$, before
reionization, which is almost completely shut down at that time.
{\textsc{thesan}} continues forming stars in these low mass haloes at a roughly
constant rate until $z=5.5$, at which time star formation in haloes of $M_{\rm
H} < 10^8$ M$_\odot$ is slowly shut down as they succumb to the strong
interstellar radiation field. Notably, haloes with $10^8 < M_{\rm H} / {\rm
M}_\odot <10^9$ do not show such a strong reduction in their global (roughly
constant) star formation rates, as shown in Fig. \ref{fig:stellarbirth}.

These drastic differences in star formation in low mass haloes also change the
types of stars that are born, as stars are created in very different environments.
Fig. \ref{fig:metallicity} shows the metallicities of all stars as a function
of time for our two models. Stars have been given a minimal metallicity of
$Z_* = 10^{-7.5}$, though almost all gas particles below this threshold are
fully primordial and are labeled as such.

All star formation in primordial and low metallicity ($Z_* \lessapprox 10^{-6}$)
gas is shut down at the onset of the Uniform UVB at $z=10$. At the same time, there is
increased star formation before this time in high metallicity gas (which has been
enriched by this first generation of stars). By contrast, the {\textsc{thesan}} model shows
continual low metallicity, and primordial, star formation across cosmic time down
to the end of the simulation at $z=5$.

Overplotted is an extremely simplistic stellar population model. Pop III stars
are chosen as those that have metal mass fraction $Z_* < 10^{-7}$, and Pop II
stars have $10^{-7} < Z_* < 0.1 Z_\odot$, where here we take $Z_\odot= 0.134$.
All other stars are classified as Pop I stars. We note that here we assign the
entire stellar particle (and hence population) to each category, which is 
certainly an overestimation for Pop III stars. For instance, \citet{Jaacks2018}
only assign the first 500 M$_\odot$ of each star particle to Pop III, suggesting
that the first star formation event sampled within the entire population would
likely enrich neighbouring gas, preventing Pop III formation \citep[see
also][]{Smith2015}. We also do not model such stars in-situ, and hence do not
apply IMF corrections and the associated differences to feedback efficiency.
Nevertheless, it is useful to bin these stars into the various categories to
consider potential differences that various reionization models can create.

We see that our Pop III star formation rates are roughly $\sim 0.5$ dex higher
in the Uniform UVB simulation at high redshift, before reionization, but that
these drop to zero (implying no more conversion of primordial gas to stars) at
that time. {\textsc{thesan}}, by contrast, can continue forming these stars down to $z=5$
(the end of the line is due to a lack of sampling points at later times, not
because of the end of Pop III star formation). This is consistent with results
from \citet{Xu2016} who also find that Pop III star formation can continue in
low mass ($M_{\rm H} \sim 10^8$ M$_\odot$) haloes until late times ($z\sim7$ in
their simulations). \citet{Wise2012} and \citet{Sarmento2022} also find that Pop
III star formation should continue until late times at a consistent rate of
$10^{-4}$ M$_\odot$ yr$^{-1}$ cMpc$^{-3}$, which is $\sim 0.5$ dex lower than
ours (though, as noted above, we overestimate Pop III star formation rates
here).

Reionization also impacts the tradeoff between Pop II and Pop I star formation
at high redshift, with equivalence between Pop II and Pop III reached at
$z\approx 8$, rather than $z\approx 5$, in a universe with a Uniform UVB
rather than a full reionization model. These differences hence have a significant
potential impact when comparing expected SEDs of high redshift galaxies to
those that are observed.

\section{Conclusions}
\label{sec:conclusions}

In this paper we have explored two reionization models: a spatially-uniform, but
time varying, UV background that is commonly used in cosmological galaxy
formation simulations, and a fully self-consistent reionization model employing
the {\textsc{Arepo-RT}} code and the {\textsc{thesan}} model. Our two models reionize at
different speeds, with the Uniform UVB model reionizing instantaneously at
$z=10$, but {\textsc{thesan}} taking down to $z \approx 5$ to have a global mass-weighted
{\textsc{Hi}} fraction of $\approx 0$. We have used two high resolution volumes, with
target baryonic masses of $\sim 10^4$ M$_\odot$ (L4N512), and $\sim 10^5$
M$_\odot$ (L8N512) to investigate the impact of these two different reionization
models on galaxy formation in the early universe. We found significant
differences at low galaxy masses, with haloes of $M_{\rm H} < 10^9$~M$_\odot$
significantly impacted by the choice of reionization model, whereas the
properties of higher-mass galaxies were relatively indifferent to this choice.
Specifically, we found that:
\begin{itemize}
    \item As expected, the Uniform UVB model washes out a significant fraction
          of the cold, dense, structures at $z > 5$, before the entire volume
          reionizes with the {\textsc{thesan}} model. In Fig. \ref{fig:bigplot} we show
          the diversity of the IGM at early times that is under-represented
          in a Uniform UVB model. Both the temperature and density of the IGM
          show significant differences between both models, as demonstrated in
          Fig. \ref{fig:combinedboxes} and Fig. \ref{fig:rhot}.
    \item Our models were shown to match well with observational constraints
          on the galaxy stellar mass function from \citet{Song2016} and
          \citet{Bhatawdekar2019} (Fig. \ref{fig:smf}). For the UVLF, we showed
          that the inclusion of a slower reionization model (i.e. {\textsc{thesan}} v.s. a
          uniform UVB) can change the position of the turnover, with galaxies of
          a $M_{\rm 1500} \sim -10$ having a whole dex higher abundance in a
          {\textsc{thesan}} model at $z=6$ than in a Uniform UVB model (Fig.
          \ref{fig:uvlf}).
    \item Differences in the UVLF are caused by differences in both the occupation
          fraction of galaxies (the fraction of haloes of a given mass
          containing galaxies) and availability of star forming gas within
          individual haloes. Haloes of mass $M_{\rm H} \approx 10^8$ M$_\odot$
          show significantly higher occupation fractions (and higher
          stellar-to-halo mass ratios) down to $z=5$ under the {\textsc{thesan}}
          reionization model (Fig. \ref{fig:smhm}). There are also significant
          differences in the baryon mass ratio ($M_{\rm b} / M_{\rm H}$, $\sim
          1$ dex at $M_{\rm H} \approx 10^8$ M$_\odot$), indicating that the
          Uniform UVB is driving gas out of low mass haloes ($M_{\rm H} < 10^9$
          M$_\odot$) post-reionization (Fig.  \ref{fig:baryonmass}).
    \item These differences in abundance of both stars and gas also affect
          the morphologies of galaxies, with {\textsc{thesan}} galaxies $\sim
          0.1-0.2$ dex larger in their physical stellar extent (Fig.
          \ref{fig:sizes}), and a factor $\sim 2$ lower in stellar metallicity
          (Fig.  \ref{fig:stellarmetallicity}).
    \item Fundamentally, the differences in properties of galaxies between reionization
          models are driven by differences in the behaviour of gas in haloes
          with virial temperatures $T_{\rm vir} < 10^4$ K ($V_{\rm circ} \approx
          30$ km s$^{-1}$). As gas is photoheated by external radiation fields,
          it becomes unbound in these haloes, and hence in simulations that
          allow haloes to be fully self-shielded until late times can allow them
          to be occupied by gas for much longer (Fig. \ref{fig:cgm}). Notably,
          this means that such processes will not impact simulations that do not
          have the resolution to model these haloes (typically those with
          $m_{\rm b} > 10^6$ M$_\odot$, like EAGLE or TNG-100), but that higher
          resolution simulations (e.g. the zooms from the FIRE suite,
          \citep{Ma2019}, or TNG-50 \citep{Pillepich2019,Nelson2019b}) are missing an
          important contribution to the distribution of their overall energy
          budget.
    \item Changes in gas occupation lead to differences in star formation, with
          star formation in haloes $M_{\rm H} < 10^9$ M$_\odot$ shut off
          post-reionization. {\textsc{thesan}} allows haloes of this mass to form stars
          until the entire volume reionizes, one of the major causes of the
          different stellar metallicities (Fig. \ref{fig:stellarbirth} and Fig.
          \ref{fig:splitsfh}). Such changes then lead to very different
          predictions for the global Pop III and Pop II star formation rates
          (Fig. \ref{fig:stellarmetallicity}), with {\textsc{thesan}} able to form Pop III
          stars down to $z \approx 5$ and the Uniform UVB model curtailing such
          pathways at $z=10$. This has clear implications for in-situ and ex-situ
          star formation contributions to early galaxies, as well as the
          processing of gas accretion that feeds in-situ star formation
          \citep[e.g.][]{Keres2009, Angles-Alcazar2017b}.
\end{itemize}
Our results present a troubling picture for the applicability of future and
present planned high-resolution simulations to early galaxy formation
\citep[e.g. FireBOX;][]{Feldmann2022}.  Without the inclusion of a
spatially-varying UV background, galaxy properties (in particular the star
formation histories of any such galaxies) will be poorly constrained.

We note that any differences seen between {\textsc{thesan}}-2 galaxies and those
in {\textsc{thesan}}-TNG-2 (i.e. with a Uniform UVB) are much smaller when
considering lower resolution ($m_{\rm b} \approx 5 \times10^6$~M$_\odot$), as
described in Appendix A of \citet{Garaldi2022}. The properties of high-redshift
galaxies simulated at lower resolution ($m_{\rm b} \gtrapprox 10^6$~M$_\odot$)
are also impacted by the effects discussed in this work; even when including a
spatially-varying UV background, they are missing out on an important channel
for early galaxy formation (haloes with $M_{\rm H} \approx 10^8$) and their
crucial influence of pre-processing and ex-situ star formation even in high mass
galaxies. We see good convergence between our models once above the $M_{\rm H}
\approx 10^8$ M$_\odot$ threshold, but below this (which is only very marginally
resolved in the dark sector, and is poorly resolved in the baryonic sector for
the L8N512 volume), significant differences can occur. This is largely due to
the artificial suppression of minihalo formation at baryonic mass resolutions
$m_{\rm b} \gtrapprox 10^{4.5}$~M$_\odot$, and represents a challenge both from
a resolution and physics perspective for ongoing galaxy formation projects. For
example, the most realistic simulations of first star and galaxy formation
require extended primordial thermochemistry networks including molecular
hydrogen formation and self-shielding of LW radiation \citep[see
e.g.][]{Greif2015}.

As such, we suggest that authors aiming to study early galaxy formation, which
is particularly crucial as we enter the \emph{JWST} era, must use resolutions
that allow them to accurately model the formation of minihaloes (i.e. $m_{\rm b}
\lessapprox 10^{4.5}$~M$_\odot$ in a typical LCDM cosmology) and must employ a
spatially-varying UV background to ensure these haloes can remain self-shielded.
We likely would have seen much closer results had we employed a spatially
inhomogeneous, semi-numerical, UVB as in e.g. \citet{Bird2022, Trac2022,
Puchwein2022}. Future work should focus on development and self-consistent
calibration of such semi-numerical models for existing galaxy formation models,
though it remains to be seen whether these semi-numerical treatments can recover
the physics seen in fully non-equilibrium simulations. The impact of our small box
volume also remains unclear. A larger volume would lead to more variation in galaxy
environment, and hence more variation in reionization histories, but would contain
higher-mass galaxies that would inevitably lead to more outside-in reionization
which is better approximated by the uniform UVB scenario.

Our work also presents observational challenges for lensing fields; if a given
region was reionized rapidly, for example a field lying near an overly massive
galaxy ($M_* > 10^9$ M$_\odot$), we expect it will have properties similar to
our Uniform UVB case, whereas relatively isolated fields should allow minihalo
formation to later times. Differences between these two reionization scenarios
should be observable within the UVLF, though this is challenging given that we
only see differences at $M_{\rm 1500} > -12$.

\section*{Acknowledgements}

The authors thank Chris Lovell, Sylvia Ploeckinger, Ewald Puchwein, and John
Wise for conversations that contributed to this work.
All of the computations were performed on the Engaging cluster
supported by the Massachusetts Institute of Technology. MV acknowledges support
through NASA ATP 19-ATP19-0019, 19-ATP19-0020, 19-ATP19-0167, and NSF grants
AST-1814053, AST-1814259, AST-1909831, AST-2007355 and AST-2107724.
AS acknowledges support for Program number \textit{HST}-HF2-51421.001-A provided
by NASA through a grant from the Space Telescope Science Institute, which is
operated by the Association of Universities for Research in Astronomy,
incorporated, under NASA contract NAS5-26555.

Software citations:
\begin{itemize}
	\item {\textsc{AREPO-RT}}: \citet{Springel2010, Kannan2019, Weinberger2020}
	\item {\textsc{Python}}: \citet{vanRossum1995}
	\item {\textsc{Matplotlib}}: \citet{Hunter2007}
	\item {\textsc{SciPy}}: \citet{Virtanen2020}
	\item {\textsc{Scikit-Learn}}: \citet{scikit-learn}
	\item {\textsc{NumPy}}: \citet{Harris2020}
	\item {\textsc{unyt}}: \citet{Goldbaum2018}
	\item {\textsc{SwiftSimIO}}: \citet{Borrow2020, Borrow2021}
	\item {\textsc{WebPlotDigitize}}: \citet{Rohatgi2022}
\end{itemize}

\section*{Data Availability}


All thesan and thesan-hr simulation data will be made publicly available in the
near future, including Ly$\alpha$ catalogues. Data will be distributed via
\url{www.thesan-project.com}. Before the public data release, data underlying
this article will be shared on reasonable request to the corresponding
author(s).



\bibliographystyle{mnras}
\bibliography{bibliography} 




\appendix
\section{Resolution Convergence}
\label{app:res}

\begin{figure}
    \centering
    \includegraphics{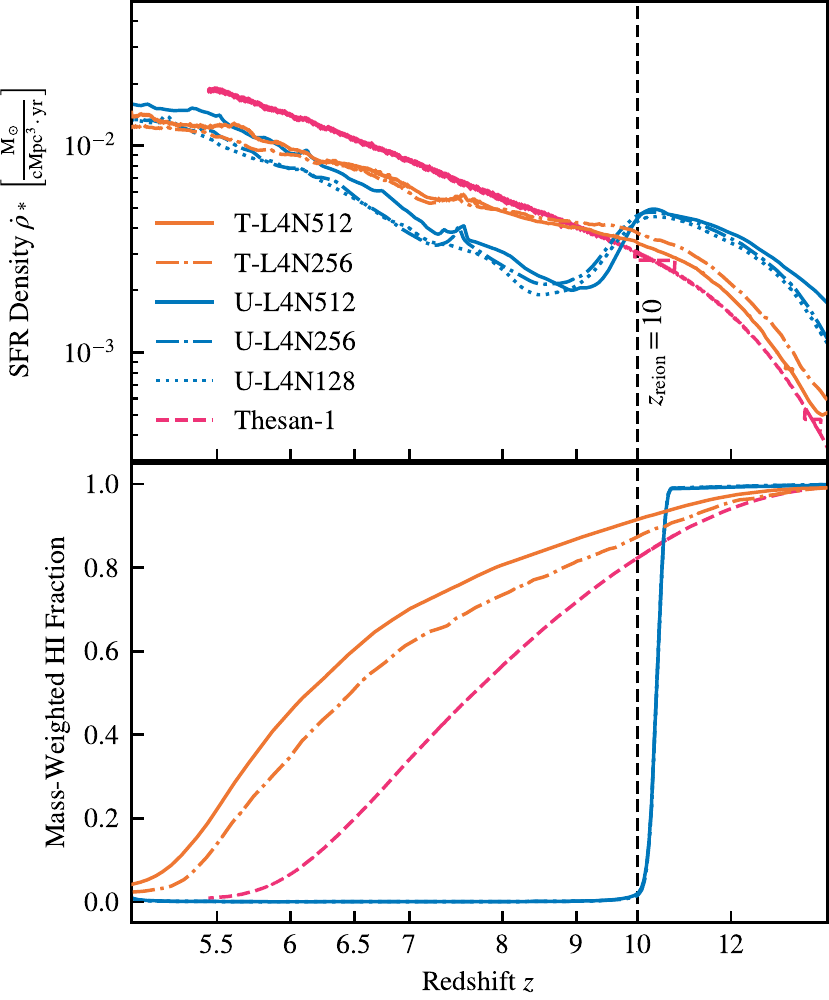}
    \caption{Analogue of Fig. \ref{fig:reionizationhistory}, which shows
    the star formation rate density of the volumes with the {\textsc{thesan}}
    model and the Uniform UVB model (line colours). The different line styles
    show different simulation resolutions, from the highest resolution used here
    to one comparable to the original {\textsc{Thesan}}-1 simulation.}
    \label{fig:hifrac_convergence}
\end{figure}

In Fig. \ref{fig:hifrac_convergence}, we demonstrate the convergence properties
of our simulations. Here, we show simulations of the same L4 volume at various
resolutions: L4N512 (as in the main text), L4N256, at the same mass resolution
as the L8N512 volume, and L4N128, which is at a resolution 8 times lower
(comparable to Thesan-1).  We see that the star formation rate density is mainly
dependent on the choice of reionization model, \emph{not} on the resolution that
the volume is simulated at. This indicates the differences between the star
formation histories of L8N512 and L4N512 in Fig. \ref{fig:reionizationhistory}
are mainly driven by the different phases and volumes than they are by
differences between baryonic resolution.

\begin{figure}
    \centering
    \includegraphics{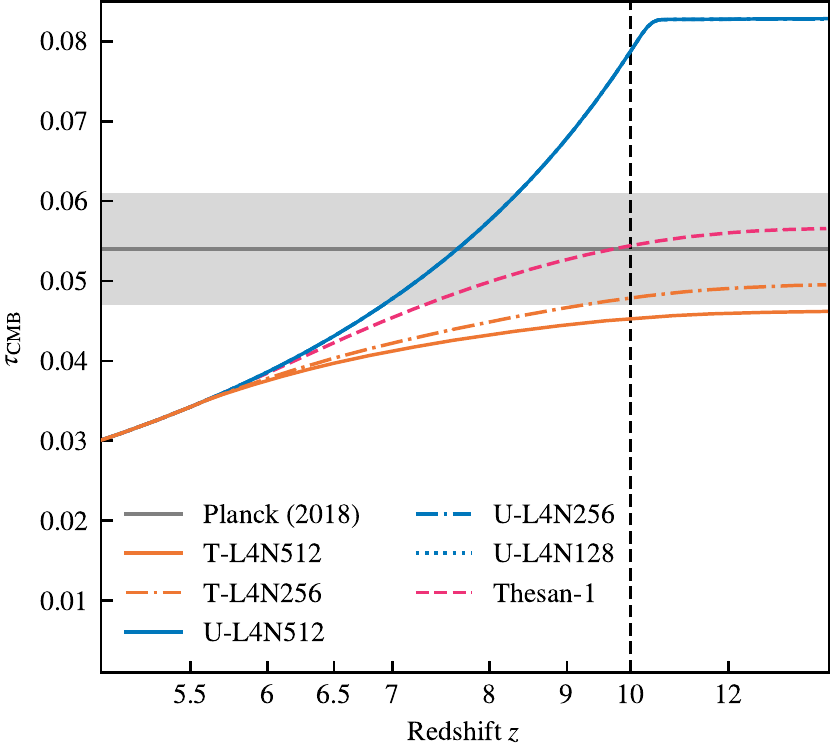}
    \caption{The optical depth to the CMB as a function of redshift, for the
    convergence volumes with a uniform UVB (blue lines) and with the standard
    {\textsc{thesan}} full radiative transfer treatment (orange lines). We show, for
    comparison, the original {\textsc{thesan}-1} simulation (pink) and the
    results from \citet{PlanckCollaboration2020}.}
    \label{fig:taucmb}
\end{figure}

In Fig. \ref{fig:taucmb} we show the optical depth to the CMB as a function of
redshift for our models at various resolutions. We see, as expected, that the
{\textsc{thesan-hr}} volumes (orange lines) under-shoot the
\citet{PlanckCollaboration2020} results due to their systematically late
reionization, originating from the lack of bright sources in these small
volumes. We see a small difference between resolution levels that is entirely
consistent with the differences seen in the star formation history (see Fig.
\ref{fig:hifrac_convergence}) and associated small change in the ionising photon
budget. As shown in \citet{Kannan2022} the {\textsc{thesan}-1} volume matches
the available Planck data well. Finally, the models employing a uniform UVB lead
to systematically high optical depths due to their vastly different (and
earlier) reionization histories.


\bsp	
\label{lastpage}
\end{document}